\documentclass[12pt]{article}
\usepackage{graphicx}
\def\hybrid{\topmargin 0pt      \oddsidemargin 0pt
        \headheight 0pt \headsep 0pt
       \voffset-1cm
        \textwidth 6.25in       % A4 paper
       \textheight 9.5in       % A4 paper
        \marginparwidth 0.0in
        \parskip 5pt plus 1pt   \jot = 1.5ex}
\catcode`\@=11
\def\marginnote#1{}

\newcount\hour
\newcount\minute
\newtoks\amorpm
\hour=\time\divide\hour by60
\minute=\time{\multiply\hour by60 \global\advance\minute by-\hour}
\edef\standardtime{{\ifnum\hour<12 \global\amorpm={am}%
        \else\global\amorpm={pm}\advance\hour by-12 \fi
        \ifnum\hour=0 \hour=12 \fi
        \number\hour:\ifnum\minute<10 0\fi\number\minute\the\amorpm}}
\edef\militarytime{\number\hour:\ifnum\minute<10 0\fi\number\minute}

\def\draftlabel#1{{\@bsphack\if@filesw {\let\thepage\relax
   \xdef\@gtempa{\write\@auxout{\string
      \newlabel{#1}{{\@currentlabel}{\thepage}}}}}\@gtempa
   \if@nobreak \ifvmode\nobreak\fi\fi\fi\@esphack}
        \gdef\@eqnlabel{#1}}
\def\@eqnlabel{}
\def\@vacuum{}
\def\draftmarginnote#1{\marginpar{\raggedright\scriptsize\tt#1}}

\def\draftlabel#1{{\@bsphack\if@filesw {\let\thepage\relax
   \xdef\@gtempa{\write\@auxout{\string
      \newlabel{#1}{{\@currentlabel}{\thepage}}}}}\@gtempa
   \if@nobreak \ifvmode\nobreak\fi\fi\fi\@esphack}
        \gdef\@eqnlabel{#1}}
\def\@eqnlabel{}
\def\@vacuum{}
\def\draftmarginnote#1{\marginpar{\raggedright\scriptsize\tt#1}}

\def\draft{\oddsidemargin -.5truein
        \def\@oddfoot{\sl preliminary draft \hfil
        \rm\thepage\hfil\sl\today\quad\militarytime}
        \let\@evenfoot\@oddfoot \overfullrule 3pt
        \let\label=\draftlabel
        \let\marginnote=\draftmarginnote
   \def\@eqnnum{(\theequation)\rlap{\kern\marginparsep\tt\@eqnlabel}%
\global\let\@eqnlabel\@vacuum}  }

\def\numberbysection{\@addtoreset{equation}{section}
        \def\theequation{\thesection.\arabic{equation}}}

\def\underline#1{\relax\ifmmode\@@underline#1\else
        $\@@underline{\hbox{#1}}$\relax\fi}

\def\titlepage{\@restonecolfalse\if@twocolumn\@restonecoltrue\onecolumn
     \else \newpage \fi \thispagestyle{empty}\c@page\z@
        \def\thefootnote{\fnsymbol{footnote}} }

\def\endtitlepage{\if@restonecol\twocolumn \else  \fi
        \def\thefootnote{\arabic{footnote}}
        \setcounter{footnote}{0}}  %\c@footnote\z@ }
\relax

\numberbysection
\hybrid

\newfont{\Bbb}{msbm10 scaled 1\@ptsize00}
\newfont{\Bbbb}{msbm7 scaled 1\@ptsize00}
\newcommand{\CC}{\mbox{\Bbb C}}

\newcommand{\DDD}{\raise-1pt\hbox{$\mbox{\Bbbb D}$}}

\newcommand{\UUU}{\raise-1pt\hbox{$\mbox{\Bbbb U}$}}

\newcommand{\ZZ}{\mbox{\Bbb Z}}
\newcommand{\z}{\raise-1pt\hbox{$\mbox{\Bbbb Z}$}}

\newcommand{\sss}{\raise-1pt\hbox{$\mbox{\Bbbb S}$}}

\def\beq{\begin{equation}}
\def\eeq{\end{equation}}
\def\p{\partial}

\newtheorem{lemma-definition}{Lemma-Definition}[section]

\def\res{\mathop{\hbox{res}}\limits}

\begin{document}

\begin{titlepage}

\title{Classical facets of quantum integrability}

\author{A. Zabrodin\thanks{Skolkovo Institute of Science and
Technology, 143026, Moscow, Russia and
National Research University Higher School of
Economics,
20 Myasnitskaya Ulitsa, Moscow 101000, Russia, 
and NRC ``Kurchatov institute'', Moscow, Russia;
e-mail: zabrodin@itep.ru}}

\date{January 2025}

\maketitle

\vspace{-7cm} \centerline{ \hfill ITEP-TH-06/25}\vspace{7cm}

\begin{abstract}

This paper is a review of the works devoted to understanding 
and reinterpretation of the theory 
of quantum integrable models solvable by Bethe ansatz in terms of
the theory of purely classical soliton equations.
Remarkably, studying polynomial solutions of the latter
by methods of classical soliton theory,
one is able to develop a method of solving the spectral
problem for the former which provides an alternative 
to the Bethe ansatz procedure.
Our main examples are
the generalized inhomogeneous spins chains
with twisted boundary conditions on the quantum side and
the modified Kadomtsev-Petviashvili hierarchy of 
nonlinear differential-difference equations on the classical
side. In this paper, we restrict ourselves to quantum spin 
chains with rational $GL(n)$-invariant 
$R$-matrices (of the XXX type). Also, the connection of quantum
spin chains with classical soliton equations implies 
a close interrelation between the spectral problem
for spin chains and integrable
many-body systems of classical mechanics such as Calogero-Moser
and Ruijsenaars-Scheider models, which is known as the 
quantum-classical duality. Revisiting this topic, we suggest
a simpler and more instructive proof of this kind of duality.

\end{abstract}

\end{titlepage}

\vspace{5mm}

%

%\newpage
\tableofcontents

\vspace{5mm}

\section{Introduction}

Integrable models of different types, from mechanical systems
with a finite number of degrees of freedom to models of field
theory, play an outstanding role in modern mathematical physics.
As a rule, they have important applications to physical problems 
as well as deep and beautiful mathematical structures 
and symmetries underlying their integrability. That is why
integrable models are interesting from both physical and
mathematical points of view.

An intriguing phenomenon in the world of integrable
models, which already has been observed in many examples,
is that the models belonging to very different
classes (for example, such as mechanical systems, spin chains, nonlinear
partial differential equations) have nontrivial
hidden interrelations. One such interrelation
is the existence of various {\it dualities} connecting 
differently looking models. Another one is a rather mysterious 
appearance of classical integrable equations as exact relations
built into the structure of quantum integrable systems even 
at $\hbar \neq 0$. Moreover, such relations sometimes 
allow one to
develop alternative methods to diagonalize commuting Hamiltonians
of the quantum systems, based on purely classical theory.

The latter program was first realized in \cite{KLWZ97} 
(see also \cite{Z97,Z98}), where 
functional relations for higher conserved quantities
of generalized quantum spin chains 
\cite{Cherednik,BR90,KNS94} were interpreted as
a classical discrete dynamical system which was 
identified with the  
Hirota bilinear difference equation \cite{Hirota81} 
known in 
soliton theory since 1981. (It is often referred to as a 
fully discrete KP equation.) 
It was demonstrated that this interpretation 
allows one to solve the spectral problem for spin chains 
using methods of classical soliton theory.
Later, in \cite{KSZ08,Z08} (see also \cite{Hegedus10}
for a further development) this approach was extended to 
generalized spin chains with superalgebra 
symmetries (super-spin chains or 
``graded magnets'' introduced by in \cite{Kulish}). 
This extension was based on the functional relations 
for higher transfer matrices of the super-spin chains
established in \cite{Tsuboi97,Tsuboi97a}.

A deeper understanding of the connection between quantum
spin chains and soliton equations was achieved in the works
\cite{AKLTZ13}--\cite{Z15}, where continuous flows
parametrized by infinitely many continuous 
parameters ${\bf t}=\{t_1, t_2, t_3, \ldots \}$ 
(``time variables'') were defined in the space of
commuting conserved quantities 
of a quantum spin chain. 
The dynamics in the times ${\bf t}$ was then identified with
the modified Kadomtsev-Petviashvili (mKP) hierarchy of nonlinear
integrable equations. In this paper we give a review of the works
\cite{AKLTZ13}--\cite{Z15}, refining some arguments from them
and making some statements more precise. 

We consider 
inhomogeneous $GL(n)$-invariant
spin chains with twisted boundary conditions. 
The commuting transfer matrices
${\sf T}_{\lambda}(x)$ depending 
on the spectral parameter $x\in \CC$
and the Young diagram $\lambda$ are constructed with the help
of the $R$-matrix
\beq\label{i2}
{\sf R}^{\lambda}_{01}(x)=x{\sf I}+\eta \sum_{a,b=1}^n
\pi_{\lambda}({\bf e}_{ab})^{(0)}\otimes e_{ba}^{(1)}
\eeq
acting in the tensor product of two linear 
spaces $V_0\times V_1$,
where ${\sf I}$ is the unity matrix, ${\bf e}_{ab}$
are generators of the universal enveloping algebra
$U(gl_n)$, $\pi_{\lambda}$ is the 
finite-dimensional irreducible representation of  
$U(gl_n)$
associated with the Young diagram $\lambda$ and $e_{ab}$ is
${\bf e}_{ab}$ in the $n$-dimensional vector representation.
The transfer matrix ${\sf T}_{\lambda}(x)$ is defined as
\beq\label{i3}
{\sf T}_{\lambda}(x)=\mbox{tr}_{0}
\Bigl ({\sf R}^{\lambda}_{01}(x-x_1)
{\sf R}^{\lambda}_{02}(x-x_2)
\ldots {\sf R}^{\lambda}_{0N}(x-x_N)\, \pi_{\lambda}({\bf g}_0)\Bigr ),
\eeq
where ${\bf g}=\mbox{diag} (p_1, \ldots , p_n)$ 
is the diagonal twist matrix and $x_i$ are arbitrary 
(in general, complex) numbers
called inhomogeneity parameters (we assume that all of them are distinct).
The standard method of simultaneous diagonalization of the 
transfer matrices is the Bethe ansatz,
coordinate or algebraic \cite{Gaudin}--\cite{Slavnov}.
For $n>2$ this technique is called the nested Bethe ansatz.
As a result, 
the eigenvalues of the transfer matrices are expressed through
a set of auxiliary quantities (Bethe roots) satisfying 
a system of algebraic equations called (nested) Bethe equations
(see \cite{KL83,BR08}).

Let ${\sf T}(x, {\bf t})$ be the following
generating function of the commuting transfer matrices:
\beq\label{i1}
{\sf T}(x, {\bf t})=\sum_{\lambda}
{\sf T}_{\lambda}(x)s_{\lambda}({\bf t}),
\eeq
where $s_{\lambda}({\bf t})$ are Schur polynomials \cite{Macdonald}
and the sum goes over all Young diagrams including the empty one
(actually, for $GL(n)$-invariant models the sum is restricted
to diagrams with not
more than $n$ non-empty rows). In \cite{AKLTZ13} 
the generating function (\ref{i1}) 
was called {\it the master $T$-operator} (see also \cite{KLT12},
where it was introduced in an implicit form).
Its main property, proved in \cite{AKLTZ13}, is that it 
satisfies the bilinear equations for the tau-function of the
mKP hierarchy, with $x$ playing the role of the ``zeroth time'' variable.
The generating bilinear equation has the form \cite{JM83,DJKM83}
\beq\label{master11b}
\begin{array}{l}
\displaystyle{
\oint_{C_{\infty}}z^{(x-x')/\eta}
e^{\xi ({\bf t}-{\bf t}', z)}
{\sf T}\left (x; {\bf t}-[z^{-1}]\right ){\sf T}\left (x' ; 
{\bf t}'+[z^{-1}]\right )dz}
\\ \\
\phantom{aaaaaaaaaaaaaaaaaaaaaa}\displaystyle{ =
\delta_{(x-x')/\eta , -1}
{\sf T}(x+\eta ;{\bf t})
{\sf T}(x'-\eta ;{\bf t}').}
\end{array}
\eeq
It is valid for all ${\bf t}, {\bf t}'$
and $x, x'$ such that $(x-x')/\eta \in \ZZ$ and
$(x-x')/\eta \geq -1$. 
In (\ref{master11b}) we use the standard notation
$\displaystyle{
\xi ({\bf t}, z)=\sum_{k\geq 1}t_kz^k}$,
${\bf t}\pm [z^{-1}]=\Bigl \{ t_1 \pm z^{-1}, t_2\pm \frac{1}{2}\, z^{-2},
t_3\pm \frac{1}{3}\, z^{-3}, \ldots \Bigr \}$.
The integration contour  
$C_{\infty}$ is a big circle of radius $R\to \infty$ 
Therefore, each eigenvalue $T(x, {\bf t})$ is a solution to the 
mKP hierarchy.

A further analysis shows that the objects arising in the process 
of solving the quantum model using the algebraic 
Bethe ansatz technique and specific for the quantum theory
have their counterparts in the classical theory of soliton
equations. 
For example, 
the classical facet of the nested Bethe ansatz procedure is a chain
of B\"acklund transformations (which we call 
the ``undressing chain'') that gradually ``undress''
a solution of the mKP hierarchy to the trivial one. Each step
of this chain corresponds to a level of the nested Bethe ansatz. 
The eigenvalues of the Baxter's $Q$-operators are nothing else than
tau-functions arising at different steps of the undressing chain.
Then the Bethe equations for their roots acquire a nice
interpretation of equations of motion for the Ruijsenaars-Schneider
system of particles in discrete time (each step of the discrete
time corresponds to a level of the Bethe ansatz). 
Further, the non-commutative generating function
for transfer matrices in fundamental representations 
becomes, on the classical side, 
the so-called wave operator ${\bf W}(x)$ 
(known also as the dressing
operator). In general, it is an infinite series in inverse
powers of the shift operator $e^{\eta \p_x}$ but a characteristic
feature of solutions relevant to the $GL(n)$-invariant 
quantum spin chains is that
this series truncates at the $n$-th term, and this operator
admits a factorization as a finite product of first order difference
operators. This factorization plays a key role in the algebraic
Bethe ansatz solution since the coefficients of the first order
operators in each factor are constructed from eigenvalues of the
$Q$-operators $Q_k(x)$, $k=1, \ldots , n-1$, so eigenvalues of the
transfer matrices can be expressed through the $Q_k$'s whose roots
are subject to the system of Bethe equations.

The mKP hierarchy has a lot of solutions of very different nature.
So, to make the correspondence with the classical theory
complete, one should characterize the class of 
solutions corresponding to
eigenvalues of the master $T$-operator. 
Since its matrix elements are polynomial in $x$ of degree $N$
and the master $T$-operators for all $x$ can be simultaneously
diagonalized, the same is true for the eigenvalues, i.e.,
the eigenvalues $T(x, {\bf t})$ are polynomials in $x$ of degree
$N$ whose roots depend on the times $t_i$:
\beq\label{master13b}
T(x; {\bf t})=C({\bf t})
\prod_{k=1}^{N}\left (x-x_k({\bf t})\right ).
\eeq
The dynamics of the zeros of polynomial tau-functions 
is a well known subject
in the theory of integrable nonlinear partial differential equations.
In the works of Krichever and others 
(see \cite{AMM77}--\cite{Z19}) it was found that  
these dynamics are described 
by equations of motion of integrable 
many-body systems of particles of the Calogero-Moser 
and Ruijsenaars-Schneider type. 
In particular, the dynamics of zeros of the tau-function of 
the mKP hierarchy
of the form (\ref{master13b}) in the times $t_k$ 
coincide with the dynamics of the Ruijsenaars-Schneider
system of particles \cite{RS86} (which is also known as a relativistic 
deformation of the Calogero-Moser system \cite{Calogero75,Moser75},
the parameter $\eta$ playing the role of the inverse velocity
of light)
with respect to the 
$k$-th Hamiltonian flow (for all $k$ 
this result was obtained in \cite{Iliev07}).
The Ruijsenaars-Schneider system admits the Lax representation
with the Lax matrix
\beq\label{Lax}
L_{ij}=\frac{\p_{t_1}x_i}{x_i-x_j -\eta}.
\eeq
The time evolution is an isospectral transformation of $L$ 
and the characteristic polynomial $\det (zI -L)$ is a generating
function of conserved quantities.

From this it follows a nontrivial 
connection between the generalized inhomogeneous
quantum spin chains solvable by the algebraic Bethe ansatz and classical integrable 
many-body systems of the Calogero-Moser type. 
This connection is called the quantum-classical duality.
To formulate it, we need to consider the transfer matrix
${\sf T}_{(1)}(x)$ (\ref{i3}) 
corresponding to the vector representation
(for which $\lambda$ is one box) and define quantum 
commuting Hamiltonians of the inhomogeneous spin chain as
\beq\label{i4}
{\bf H}_i =\eta^{-1}\res_{x=x_i}\left (
\frac{{\sf T}_{(1)}(x)}{\prod\limits_{k=1}^N
(x-x_k)}\right ).
\eeq
Then the quantum-classical duality states that
the eigenvalues $H_i$ of the ${\bf H}_i$'s are given by velocities 
$\dot x_i(0)=\p_{t_1}x_i \Bigr |_{{\bf t}=0}$
of the classical Ruijsenaars-Schneider particles at ${\bf t}=0$:
\beq\label{i5}
H_i=-\eta^{-1}\dot x_i(0),
\eeq
which have to be found from the condition that the spectrum 
of the Lax matrix $L$ has the form
\beq\label{qc3a}
\mbox{Spec}\, L =
\Bigl (\underbrace{p_1, \ldots , p_1}_{M_1}, \, 
\underbrace{p_2, \ldots , p_2}_{M_2}, \,
\ldots , \, \underbrace{p_n, \ldots , p_n}_{M_n}\Bigr ),
\eeq
where $M_a$ are the eigenvalues of 
the weight operators $\displaystyle{{\bf M}_a=
\sum_{k=1}^N \pi_{(1)}({\bf e}^{(k)}_{aa})}$ on the common 
eigenstate with the transfer matrix and
$p_i$ are the twist parameters (elements of the diagonal twist
matrix ${\bf g}$).

A direct proof of this statement was given
in \cite{GZZ14} for models with rational dependence
on the spectral parameter $x$ and in \cite{BLZZ16} for models
with trigonometric $R$-matrices. By other reasoning and in 
another context, this kind of duality was discussed in
\cite{GK13} (see also \cite{GK95,MTV12}, where 
similar phenomena were observed in 
some simpler particular and limiting 
cases). However, the proof given in \cite{GZZ14}
was essentially based on the nested Bethe equations
and was technically involved. In this paper, we suggest another,
much simpler proof, which avoids any explicit use of Bethe equations.

The structure of the paper is as follows.
In Section 2 we recall the standard definitions 
and facts from the theory of 
integrable $GL(n)$-invariant spin chains. In Section 3 
we introduce the higher transfer matrices ${\sf T}_{\lambda}(x)$
indexed by the Young diagrams $\lambda$ and recall the functional
relations for them, which have the form of the spectral parameter
dependent determinant identities of the Jacobi-Trudi type.
The master $T$-operator is introduced in Section 4.1, and in Section
4.2 the integral bilinear relation for it is given. This relation
allows one to identify eigenvalues of the master $T$-operator with
tau-functions of the mKP hierarchy. Section 5 is devoted to
the basic facts related to the mKP hierarchy. In particular, we recall
the construction of the wave operator ${\bf W}(x, {\bf t})$, the 
wave function $\psi (x, {\bf t};z)$ and its adjoint 
$\psi^*(x,{\bf t};z)$. In Section 6 we study (quasi)polynomial
solutions to the mKP hierarchy. They are constructed using the
approach suggested by Krichever, which is based on the conditions
(\ref{rat2a}) imposed on the wave function as a function 
of the spectral parameter $z$. Also, in this section the 
undressing and dressing B\"acklund
transformations are considered, and a chain of the B\"acklund
transformations is introduced. The index that labels 
steps of the chain can be regarded as a discrete time variable.
The dynamics of zeros of the tau-functions in this discrete
time is the Ruijsenaars-Schneider system in discrete
time. Remarkably, its equations of motion 
are the nested Bethe equations. In Section 7 we present a detailed
identification of the objects from the algebraic Bethe ansatz 
with the ones from the classical theory of the mKP hierarchy.
In Section 8 the connection with the classical Ruijsenaars-Schneider
many-body system is established, and the quantum-classical 
duality is discussed. Finally, in Section 9 we make a few concluding
remarks and mention some open problems.

\section{Generalized spin chains}

\subsection{$GL(n)$-invariant $R$-matrices}

Integrable $GL(n)$-invariant spin chains and vertex models are constructed 
by means of $R$-matrices depending on a spectral parameter $x\in \CC$. 
Let $V= \CC^n$ be the $n$-dimensional 
linear space of vector representation of the group $GL(n)$.
The simplest $GL(n)$-invariant $R$-matrix ${\sf R}(x)$ 
is a linear operator on
$V\otimes V$. It can be represented as a matrix 
of size $n^2\! \times \! n^2$ for any $n\geq 2$.

Let $V_1, V_2, V_3$ be three copies of the space $V$. 
By ${\sf R}_{12}(x)$ we denote the $R$-matrix that acts 
nontrivially on $V_1\otimes V_2$ and trivially in $V_3$, etc.
The $R$-matrix is required to satisfy the Yang-Baxter equation
\beq\label{gln4a}
{\sf R}_{12}(x-x') {\sf R}_{13}(x) {\sf R}_{23}(x')=
{\sf R}_{23}(x') {\sf R}_{13}(x) {\sf R}_{12}(x-x'),
\eeq
where the both sides are linear operators 
on the space $V_1\otimes V_2 \otimes V_3$.
The $GL(n)$-invariance means that the ${\sf R}(x)$ commutes with
${\bf g}\otimes {\bf g}$ for any ${\bf g}\in GL(n)$, i.e.,
\beq\label{gln2}
{\bf g}\otimes {\bf g} \, {\sf R}(x )={\sf R}(x )\, {\bf g}\otimes {\bf g}\qquad
\mbox{or} \qquad {\bf g}_1 {\bf g}_2 {\sf R}_{12}(x )=
{\sf R}_{12}(x ){\bf g}_1 {\bf g}_2.
\eeq
 
It is known that there are $GL(n)$-invariant 
solutions of the Yang-Baxter equation
of size $n^2\! \times \! n^2$ for any $n\geq 2$ having polynomial 
dependence on the spectral parameter. 
These $R$-matrices have the form
\beq\label{gln1}
{\sf R}(x)=x{\sf I}+\eta 
\sum_{a,b=1}^n e_{ab}\otimes e_{ba},
\eeq
where ${\sf I}$ is the unity matrix 
in $V\otimes V$, $e_{ab}$ are
elementary $n\! \times \! n$ matrices 
with 1 at the place $ab$ and 0 otherwise and
$\eta$ 
is a parameter\footnote{This parameter can be 
absorbed in $x$ but we prefer to keep 
it alive in order to prepare for the limit to the Gaudin 
model which is $\eta \to 0$.}.
Note that 
$$
{\sf P}=\sum_{a,b=1}^n e_{ab}\otimes e_{ba}
$$
is the 
permutation operator in 
$V\otimes V$, so the $R$-matrix can be written as
${\sf R}(x)=x{\sf I}+\eta {\sf P}$.
In what follows, the permutation
operator in $V_i\otimes V_j$ is denoted by ${\sf P}_{ij}$. 

\subsection{Transfer matrices}

The generalized spin chain or vertex model based on 
the
$GL(n)$-invariant $R$-matrix is introduced via the
transfer matrix which is the generating function of conserved 
quantities (a family of commuting operators). The quantum space 
of the model on $N$ sites is ${\cal V}=
\otimes_{i=1}^NV_i$. We assume that $N\geq n$.
Let $V_0$ be another copy of $\CC^n$ (it is called the auxiliary space)
and $x_1, x_2, \ldots , x_N$ be arbitrary parameters (in what follows 
we assume that they are all distinct).
The inhomogeneous model with periodic boundary conditions 
is defined by means of the
transfer matrix  
$$
{\sf T}(x)=\mbox{tr}_0\Bigl ({\sf R}_{01}(x-x_1)
{\sf R}_{02}(x-x_2)\ldots {\sf R}_{0N}(x-x_N)\Bigr ),
$$
where the trace is taken in the auxiliary space $V_0$. The transfer 
matrix is a linear operator in the space ${\cal V}$.
The Yang-Baxter equation for the $R$-matrix 
guarantees that the transfer matrices 
commute at different values of the spectral parameter:
$[{\sf T}(x), \, {\sf T}(x')]=0.$ 

More generally, one can also consider the chain 
with quasiperiodic (twisted) boundary conditions
inserting under the trace a group element
${\bf g}\in GL(n)$ (twist), which for simplicity 
we assume to be diagonal (${\bf g}=\mbox{diag}(g_1, g_2, \ldots , g_n)$):
\beq\label{master1}
{\sf T}(x)=\mbox{tr}_0\Bigl ({\sf R}_{01}(x-x_1){\sf R}_{02}(x-x_2)
\ldots {\sf R}_{0N}(x-x_N)\, {\bf g}_0\Bigr )
\eeq
The notation 
${\bf g}_0$ means that ${\bf g}$ acts in the auxiliary space (number 0).
The $GL(n)$-invariance 
(\ref{gln2}) 
implies that these transfer matrices commute at different values of the spectral parameter. 
In the homogeneous chain with periodic boundary conditions
(at $x_j=0$, ${\bf g}={\sf I}$) there exists a local Hamiltonian which is 
the logarithmic derivative of
${\sf T}(x)$ at $x=0$. 
In inhomogeneous chains local Hamiltonians 
commuting with the transfer matrix 
in general do not exist.

Matrix elements of the transfer matrix (\ref{master1}) 
are polynomials in $x$
of degree not greater than $N$. 
Let us normalize the transfer matrix in a different way, 
dividing it by the polynomial
\beq\label{phi}
\phi (x)=\prod_{j=1}^N (x-x_j).
\eeq
In this normalization the transfer matrix
\beq\label{bfT} 
{\bf T}(x)= \frac{{\sf T}(x)}{\prod\limits_{j=1}^N (x-x_j)}
\eeq
has simple poles at the points $x_j$.
Obviously, the transfer matrix ${\bf T}(x)$ is given by
\beq\label{master1a}
{\bf T}(x)=\mbox{tr}_0
\Bigl (\tilde {\sf R}_{01}(x-x_1)\tilde {\sf R}_{02}(x-x_2)
\ldots \tilde {\sf R}_{0N}(x-x_N)\, {\bf g}_0\Bigr ),
\eeq
where 
$$\displaystyle{\tilde {\sf R}(x)={\sf I}+\frac{\eta}{x}\, {\sf P}}$$
is the $R$-matrix which differs from the ${\sf R}(x)$ by a scalar factor. 
One can introduce Hamiltonians 
${\bf H}_j$ of the inhomogeneous spin chain as residues 
of ${\bf T}(x)$ at the poles:
\beq\label{master2}
{\bf T}(x)=\mbox{tr}\, {\bf g}+\sum_{j=1}^N \frac{\eta {\bf H}_j}{x-x_j}.
\eeq
These operators commute with each other. However, they are non-local. 
Their explicit form is
as follows:
\beq\label{master2a}
{\bf H}_i=\tilde {\sf R}_{i\, i\! -\! 1}(x_i-x_{i-1})\ldots
\tilde {\sf R}_{i1}(x_i-x_{1}){\bf g}_i 
\tilde {\sf R}_{iN}(x_i-x_{N})\ldots
\tilde {\sf R}_{i\, i\! +\! 1}(x_i-x_{i+1}).
\eeq
Comparing the expansions as $x\to \infty$ of (\ref{master2}) and
$$
\begin{array}{lll}
{\bf T}(x)&=&\displaystyle{\mbox{tr}_0\left [
\Bigl ({\sf I}+\frac{\eta {\sf P}_{01}}{x-x_1}\Bigr ) \ldots
\Bigl ({\sf I}+\frac{\eta {\sf P}_{0N}}{x-x_N}\Bigr )\, {\bf g}_0\right ]
}
\\ &&\\
&=&\displaystyle{\mbox{tr}\, {\bf g}\cdot {\sf I} +\frac{\eta}{x}\sum_{i=1}^N 
\mbox{tr}_0 \Bigl ({\sf P}_{0i}{\bf g}_0\Bigr ) +\ldots 
=\mbox{tr}\, {\bf g}\cdot {\sf I} +\frac{\eta}{x}
\sum_{i=1}^N {\bf g}_i+\ldots ,}
\end{array}
$$
we get the following ``sum rule'':
$$
\sum_{i=1}^N {\bf H}_i = \sum_{j=1}^N {\bf g}_j,
$$
where both sides are operators on ${\cal V}=V_1 \otimes V_2 
\otimes \ldots \otimes V_N$ and ${\bf g}_j$ acts as ${\bf g}$ in $V_j$
and trivially in the other tensor factors.

Let us mention the limit of this construction as
$\eta \to 0$, which is of its own interest. In this limit the generalized 
spin chain becomes the Gaudin model. 
Set ${\bf g}=e^{\eta {\bf h}}$, then in the limit $\eta \to 0$ we have
${\bf H}_i =1+\eta {\bf H}^G_i +O(\eta^2)$, where
$$
 {\bf H}^G_i ={\bf h}_i +\sum_{j\neq i}\frac{{\sf P}_{ij}}{x_i-x_j}
$$
are commuting Gaudin Hamiltonians of the Gaudin model. 

Let $e_{ab}^{(j)}$ be the operator in ${\cal V}$
that acts as $e_{ab}$ in $V_j$
and trivially in the other tensor factors.
Consider the
operators
\beq\label{eig0}
{\bf M}_a=\sum_{j=1}^N e_{aa}^{(j)}, \qquad a=1, \ldots , n
\eeq
which are sometimes called weight operators\footnote{In the case
$n=2$ which corresponds to the XXX spin chain
with spins $\frac{1}{2}$ the operator $\frac{1}{2}({\bf M}_1-{\bf M}_2)$
is the operator of the $z$-projection of the total spin.}. 
It is easy to see that they
commute with the transfer matrix and among themselves
and can be simultaneously diagonalized. Therefore, one can find
eigenstates of the transfer matrix which are simultaneously 
eigenstates of the operators
${\bf M}_a$ with eigenvalues $M_a$. Note that
$\displaystyle{\sum_{a=1}^n {\bf M}_a =N \, {\sf I}}$,
so $\displaystyle{\sum_{a=1}^n M_a =N}$.

Let us present the result of diagonalization of the transfer matrix 
${\bf T}(x)$ by means of the algebraic (nested) Bethe ansatz. 
We give it here without derivation (see \cite{KL83,BR08} for details).
The eigenvalues $\Lambda (x)$ of ${\bf T}(x)$ are given by
\beq\label{eig1}
\Lambda (x)=
\sum_{b=1}^{n}g_b \prod_{\gamma =1}^{{\cal N}_{b-1}}
\frac{x-w_{\gamma}^{(b-1)}+\eta}{x-w_{\gamma}^{(b-1)}}
\prod_{\beta =1}^{{\cal N}_{b}}
\frac{x-w_{\beta}^{(b)}-\eta}{x-w_{\beta}^{(b)}},
\eeq
where ${\cal N}_0=N$, ${\cal N}_0\geq {\cal N}_1 
\geq {\cal N}_2 \geq \ldots \geq {\cal N}_{n-1}\geq 0$ 
are non-negative integers, 
${\cal N}_n=0$, 
$w_{\gamma}^{(0)}=x_{\gamma}$ and the sets of Bethe roots
$\{w_{\beta}^{(b)}\}_{\beta =1}^{{\cal N}_b}$ satisfy the 
system of {\it nested Bethe ansatz
equations}
\beq\label{eig2}
g_b \prod_{\gamma =1}^{{\cal N}_{b-1}}\frac{w_{\alpha}^{(b)}-
w_{\gamma}^{(b-1)}+\eta}{w_{\alpha}^{(b)}-w_{\gamma}^{(b-1)}}=
g_{b+1}\prod_{\gamma \neq \alpha}^{{\cal N}_b}
\frac{w_{\alpha}^{(b)}-w_{\gamma}^{(b)}
+\eta}{w_{\alpha}^{(b)}-w_{\gamma}^{(b)}-\eta}
\prod_{\beta =1}^{{\cal N}_{b+1}}\frac{w_{\alpha}^{(b)}-w_{\beta}^{(b+1)}
-\eta}{w_{\alpha}^{(b)}-w_{\beta}^{(b+1)}}.
\eeq
Here $b=1, \ldots , n-1$, $\alpha = 1, \ldots , {\cal N}_b$. 
The numbers ${\cal N}_a$ are such that
${\cal N}_{a-1}-{\cal N}_a=M_{n-a+1}$, 
$a=1, \ldots , n$, where $M_a$ are eigenvalues of the 
operators ${\bf M}_a$. 
The total number of equations
in the system is ${\cal N}_1+\ldots +{\cal N}_{n-1}$. As it follows from 
(\ref{master2}), (\ref{eig1}), the eigenvalues 
$H_i$ of the Hamiltonians ${\bf H}_i$
are given by
\beq\label{eig3}
H_i=g_1\prod_{k\neq i}^N 
\frac{x_i-x_k+\eta}{x_i-x_k}\prod_{\gamma =1}^{{\cal N}_1}
\frac{x_i-w_{\gamma}^{(1)}-\eta}{x_i-w_{\gamma}^{(1)}}.
\eeq
In Section \ref{section:dressing} we will obtain the system 
of Bethe equations in the context of the mKP hierarchy.

\section{Transfer matrices as generalized characters}

For $GL(n)$-invariant models with $n>2$ 
the algebra of commuting operators
is larger than the one generated by the 
Hamiltonians ${\bf H}_j$. 
In fact, there exists a larger family of
(more general) transfer matrices commuting with 
${\sf T}(x)$.

To proceed, we need some information about 
representations of the group
$GL(n)$ and the universal enveloping algebra
$U(gl_n)$ which has generators ${\bf e}_{ab}$ 
with the commutation relations
\beq\label{ee}
{\bf e}_{ab}{\bf e}_{a'b'}-{\bf e}_{a'b'}{\bf e}_{ab}=
\delta_{a'b}{\bf e}_{ab'}-\delta_{ab'}{\bf e}_{a'b}.
\eeq
Finite-dimensional irreducible representations
$\pi_{\lambda}$ of $U(gl_n)$ 
are characterized by the highest weight
$\lambda = (\lambda_1, \lambda_2, \ldots , \lambda_n)$, where
$\lambda_i$ are non-negative integer numbers such that
$\lambda_1 \geq \lambda_2 \geq \ldots \geq \lambda_n \geq 0$.
The set of numbers $\lambda_i$ can be identified with the Young diagram
$\lambda$, or, equivalently, with the partition of 
$|\lambda |=\sum_i \lambda_i$. Let 
$V_{\lambda}$ be the representation space of
$\pi_{\lambda}$. The vector representation
is $\pi_{(1)}({\bf e}_{ab})=e_{ab}$ with
$V_{(1)}=V=\CC^n$ (here $(1)$ is the Young diagram 
consisting of one box). The fundamental representations
correspond to one-column diagrams of height from $1$ to $n$. 
If the height is greater than $n$, the representation is trivial:
$\pi_{(1^m)}({\bf e}_{ab})=0$ for $m>n$. Moreover, the representation
is trivial for any $\lambda$
with $\ell (\lambda )>n$, where $\ell (\lambda )$ is the number
of non-empty rows of $\lambda$.

Two simple special cases are important for what follows. The empty
diagram $\lambda =\emptyset$ 
corresponds to the trivial representation: 
$\pi_{\emptyset}({\bf e}_{ab})=0$. The column of $n$ boxes corresponds
to the one-dimensional representation 
$\pi_{(1^n)}({\bf e}_{ab})=\delta_{ab}$.

\subsection{Higher transfer matrices}

We first introduce more general $GL(n)$-invariant $R$-matrices.
They act in the tensor product
$V_{\lambda}\otimes \CC^n$ and have the form
\beq\label{gln3}
{\sf R}^{\lambda}(x)=x{\sf I}+\eta \sum_{a,b}
\pi_{\lambda}({\bf e}_{ab})\otimes e_{ba}.
\eeq
The $GL(n)$-invariance means that  
$$
\pi_{\lambda}({\bf g})\otimes {\bf g} \, {\sf R}^{\lambda}(x )=
{\sf R}^{\lambda}(x )\, \pi_{\lambda}({\bf g})\otimes {\bf g}.
$$
The $R$-matrices ${\sf R}^{\lambda}(x)$ satisfy the Yang-Baxter equation
\beq\label{gln4}
{\sf R}^{\lambda \mu}_{12}(x-x') 
{\sf R}^{\lambda}_{13}(x) {\sf R}^{\mu}_{23}(x')=
{\sf R}^{\mu}_{23}(x') {\sf R}^{\lambda}_{13}(x) 
{\sf R}^{\lambda \mu}_{12}(x-x'),
\eeq
where ${\sf R}^{\lambda \mu}(x-x')$ is some 
$R$-matrix acting in the tensor product
$V_{\lambda}\otimes V_{\mu}$. Its explicit form for arbitrary 
$\lambda , \mu$ is complicated.

It is possible to construct more general 
transfer matrices acting in the same
quantum space
${\cal V}$, taking as the auxiliary space not
$\CC^n$ but the space $V_{\lambda}$ of an irreducible representation
$\pi_{\lambda}$ of the algebra $U(gl_n)$. 
Such a transfer matrix is obtained as the trace in
$V_{\lambda}$ of the product of the $R$-matrices 
(\ref{gln3}):
\beq\label{master3}
{\sf T}_{\lambda}(x)=\mbox{tr}_{V_{\lambda}}
\Bigl ({\sf R}^{\lambda}_{01}(x-x_1)
{\sf R}^{\lambda}_{02}(x-x_2)
\ldots {\sf R}^{\lambda}_{0N}(x-x_N)\, \pi_{\lambda}({\bf g}_0)\Bigr ).
\eeq
From the Yang-Baxter equation 
(\ref{gln4}) 
and $GL(n)$-invariance it follows that the transfer matrices 
${\sf T}_{\lambda}(x)$ commute for different $x$ and $\lambda$:
$$
[{\sf T}_{\lambda}(x), \, {\sf T}_{\mu}(x')]=0.
$$

As it was already mentioned,
the empty diagram $\lambda =\emptyset$ corresponds
to the trivial representation $\pi_{\emptyset}({\bf e}_{ab})=0$,
$\pi_{\emptyset}({\bf g})=1$
and we have from (\ref{master3}):
\beq\label{phi1}
{\sf T}_{\emptyset}(x)=\prod_{i=1}^N (x-x_i)\cdot {\sf I}=\phi (x){\sf I}.
\eeq
One can introduce normalized transfer matrices 
${\bf T}_{\lambda}(x)$
dividing by
${\sf T}_{\emptyset}(x)$:
$$
{\bf T}_{\lambda}(x)=\frac{{\sf T}_{\lambda}(x)}{{\sf T}_{\emptyset}(x)}.
$$
Then ${\bf T}_{\emptyset}(x)={\sf I}$ and
${\bf T}_{(1)}(x)={\bf T}(x)$ introduced in (\ref{master1a}). 

For the one-column diagram $\lambda =(1^n)$ of height $n$
we have one-dimensional representation 
$\pi_{(1^n)}({\bf e}_{ab})=\delta_{ab}$,
$\pi_{(1^n)}({\bf g})=\det {\bf g}$ and formula (\ref{master3}) yields:
\beq\label{master4a}
{\sf T}_{(1^n)}(x) =\det {\bf g} \, \phi (x+\eta )\, {\sf I}.
\eeq
In the quantum inverse scattering method this transfer matrix
has the meaning of quantum determinant of the quantum monodromy 
matrix. For diagrams such that $\ell (\lambda )>n$ 
the transfer matrices
vanish identically.

At $N=0$ the definition (\ref{master3}) yields:
\beq\label{master4}
{\bf T}^{(N=0)}_{\lambda}(x)=
\mbox{tr}_{V_{\lambda}}\pi_{\lambda}({\bf g})=\chi_{\lambda}({\bf g}),
\eeq
where $\chi_{\lambda}({\bf g})$ is the character of ${\bf g}$ in the representation $\pi_{\lambda}$.
Also, we have
$$
{\bf T}_{\lambda}(x)=\chi_{\lambda}({\bf g})\cdot {\sf I}+O(1/x), \qquad x \to \infty ,
$$
so the normalized transfer matrices can be regarded 
as a generalization of characters.

It is known that the characters are given by Schur polynomials
$s_{\lambda}$ of the eigenvalues
$g_i$ of the matrix ${\bf g}$:
$$
\chi_{\lambda}({\bf g})=s_{\lambda}(\{g_i\})=
\frac{\det_{ij}\left (g_i^{n+\lambda_j-j}
\right )}{\det_{ij}\left (g_i^{n-j}\right )}.
$$
The Schur polynomials are symmetric functions of 
$g_i$. It is often convenient to consider Schur polynomials 
$s_{\lambda}(\{\xi_i\})$, where $\{\xi_i\}$ 
is a set of some variables, 
as functions $s_{\lambda}({\bf t})$
of their power sums 
$t_k =\frac{1}{k}\sum\limits_i \xi_i^k$
(${\bf t}=\{t_1, t_2, t_3, \ldots \}$
is the set of these new variables, in general infinite). 
For example, 
$$
\begin{array}{c}
s_{\emptyset}({\bf t})=1, \quad
s_{(1)}({\bf t})=t_1, \quad 
s_{(2)}({\bf t})=\frac{1}{2}t_1^2+t_2, \quad
s_{(1^2)}({\bf t})=
\frac{1}{2}t_1^2-t_2 
\end{array}
$$
and so on. For any finite diagram $\lambda$ 
the polynomial $s_{\lambda}({\bf t})$
depends only on a finite number of $t_i$'s. 

The Schur polynomials corresponding to 
the diagrams that are rows or columns 
of the form, respectively, $\lambda =(s)$ or $\lambda=1^{a}$ 
play a specially important role. It is customary to use the special
notation for them \cite{Macdonald}:
\beq\label{he}
h_k({\bf t})=s_{(k)}({\bf t}), \qquad
e_k({\bf t})=s_{(1^k)}({\bf t}).
\eeq
The generating functions for them are as follows:
\beq\label{he1}
\begin{array}{l}
\displaystyle{
\exp \Bigl (\sum_{k\geq 1}t_k z^k\Bigr )=\sum_{k\geq 0}h_k({\bf t})z^k,}
\\ \\
\displaystyle{
\exp \Bigl (-\sum_{k\geq 1}t_k (-z)^k\Bigr )=\sum_{k\geq 0}e_k({\bf t})z^k.}
\end{array}
\eeq
After the substitution 
$t_k =\frac{1}{k}\sum\limits_{i=1}^m \xi_i^k$ these polynomials
become  symmetric functions of the variables $\xi_1, \ldots , \xi_m$. 
Note that $e_{m}(\{\xi_i\}_m)=\xi_1\ldots \xi_m$ 
and $e_{k}(\{\xi_i\}_m)=0$ if $m>k$.
From (\ref{he1}) it is clear that
\beq\label{he2}
e_k ({\bf t}) =(-1)^k h_k(-{\bf t}).
\eeq

The Schur polynomials satisfy a number of important identities. 
First, we should mention the 
Cauchy-Littlewood identity \cite{Macdonald}
\beq\label{master5}
\sum_{\lambda}s_{\lambda}({\bf t})s_{\lambda}({\bf t}')=
\exp \Bigl ( \sum_{k\geq 1}kt_k t'_k\Bigr ),
\eeq
where the sum on the left-hand side is taken 
over all Young diagrams including the empty one. 
Second, there are identities which express Schur polynomials
for general $\lambda$'s through those corresponding to 
the diagrams which are either rows or columns, i.e., through
$h_{\lambda}({\bf t})$ or $e_{\lambda}({\bf t})$:
\beq\label{master6a}
s_{\lambda}({\bf t})=
\det_{1\leq i,j\leq \lambda_1'}h_{\lambda_i-i+j}({\bf t}),
\eeq  
\beq\label{master7a}
s_{\lambda}({\bf t})=
\det_{1\leq i,j\leq \lambda_1}e_{\lambda_i'-i+j}({\bf t}).
\eeq
Here
$\lambda '$ is the diagram $\lambda$ 
transposed with respect to the main diagonal,
so that $\lambda_1', \lambda_2', \ldots$ are heights of columns of 
$\lambda$. 
Equations (\ref{master6a}), (\ref{master7a}) 
are called the Jacobi-Trudi identities.
In terms of characters, they have the form
\beq\label{master6}
\chi_{\lambda}({\bf g})=
\det_{1\leq i,j\leq \lambda_1'}\chi_{(\lambda_i-i+j)}({\bf g}),
\eeq  
\beq\label{master7}
\chi_{\lambda}({\bf g})=
\det_{1\leq i,j\leq \lambda_1}\chi_{(1^{\lambda_i'-i+j})}({\bf g}).
\eeq
Note that $\chi_{\lambda}({\bf g})=0$ if $\ell (\lambda )>n$.

The transfer matrices in symmetric or
antisymmetric representations, i.e., corresponding to 
the diagrams which are rows or columns 
of the form, respectively, $\lambda =(s)$ or $\lambda=1^{a}$ 
play a specially important role. In what follows we will use 
the special simplified notation for them:
\beq\label{master8b}
{\sf T}_s(x)={\sf T}_{(s)}(x), \qquad
{\sf T}^a(x)={\sf T}_{(1^{a})}(x).
\eeq
The analogy between transfer matrices and characters is 
further supported by the fact that 
the transfer matrices satisfy 
the following identities (functional relations),
which look similarly to the Jacobi-Trudi identities for characters: 
\beq\label{master8}
{\bf T}_{\lambda}(x)=
\det_{1\leq i,j\leq \lambda_1'}{\bf T}_{\lambda_i-i+j}(x-(j\! -\! 1)\eta ),
\eeq
\beq\label{master9}
{\bf T}_{\lambda}(x)=\det_{1\leq i,j
\leq \lambda_1}{\bf T}^{\lambda_i'-i+j}(x+(j\! -\! 1)\eta ).
\eeq
They are called the Cherednik-Bazhanov-Reshetikhin (CBR) 
identities or quantum 
Ja\-co\-bi-Tru\-di identities \cite{Cherednik,BR90}. 
For the transfer matrices in the initial 
normalization (\ref{master3}), they
look as follows:
\beq\label{master8a}
{\sf T}_{\lambda}(x)=
\left (\prod_{k=1}^{\lambda_1'-1}\phi (x-k\eta )\right )^{-1}
\det_{1\leq i,j\leq \lambda_1'}{\sf T}_{\lambda_i-i+j}
(x-(j\! -\! 1)\eta ),
\eeq
\beq\label{master9a}
{\sf T}_{\lambda}(x)=
\left (\prod_{k=1}^{\lambda_1-1}\phi (x+k\eta )
\right )^{-1}\det_{1\leq i,j
\leq \lambda_1}{\sf T}^{\lambda_i'-i+j}(x+(j\! -\! 1)\eta ).
\eeq
In fact the relations (\ref{master8a}) and 
(\ref{master9a}) are equivalent: (\ref{master9a}) follows
from (\ref{master8a}) and vice versa.
Note that since ${\sf T}_{\lambda}(x)$ are polynomials, 
the possible 
poles at $x=x_i \pm k\eta$
coming from the pre-factors in the right-hand sides,
must be cancelled by zeros of the determinants.
Like the characters, the transfer matrices ${\sf T}_{\lambda}(x)$
vanish identically if $\lambda_1'>n$ (in particular, 
${\sf T}^a(x)$ vanishes if $a>n$).

\subsection{The approach based on co-derivative}

There is an elegant way, suggested in \cite{KV08}, 
to represent the transfer matrices 
${\sf T}_{\lambda}(x)$ through matrix derivatives of the characters
$\chi_{\lambda}({\bf g})$ with respect to the twist 
matrix ${\bf g}$ (for this purpose it is 
not assumed to be diagonal). This approach is an alternative
to the fusion procedure.

Let $f({\bf g})$ be any function on the group $GL(n)$
(${\bf g}\in GL(n)$). Define the matrix derivative (which is
called co-derivative in \cite{KV08})
as follows: 
\beq\label{co1}
Df({\bf g})=\sum_{a,b}e_{ab}\frac{\p}{\p \varepsilon}f
\left (e^{\varepsilon {\bf e}_{ba}}{\bf g}
\right )\Bigr |_{\varepsilon =0}.
\eeq
According to this definition, we see that if values of 
$f$ belong to a space $W$, then values of
$Df({\bf g})$ belong to ${\rm End}(\CC^n)\otimes W$. 
For example, we have:
\beq\label{co1a}
D \det {\bf g} = \det {\bf g} \cdot {\sf I}.
\eeq

An equivalent definition in 
components is
$$
D_b^a=\sum_c g^a_c\frac{\p}{\p g^b_c},
$$
where $g^a_b$ are matrix elements of the matrix ${\bf g}\in GL(n)$ in the vector representation.
Explicitly, we have:
$$
D_b^af({\bf g})=\frac{\p}{\p 
\varepsilon}f\left (e^{\varepsilon {\bf e}_{ba}}{\bf g}
\right )\Bigr |_{\varepsilon =0}.
$$
A direct calculation of the commutator $[D_{b_2}^{a_2}, 
D_{b_1}^{a_1}]$ shows that
\beq\label{co2}
[D_{b_2}^{a_2}, D_{b_1}^{a_1}]=\delta_{a_1 b_2}D^{a_2}_{b_1}-\delta_{a_2b_1}D^{a_1}_{b_2},
\eeq
i.e., the operators $D^a_b$ have the same commutation
relations as the generators ${\bf e}_{ab}$ of the algebra $U(gl_n)$.

In the case when the co-derivatives 
act on functions with values in
$\mbox{End} (\otimes_i V_i)$, it is convenient to modify the notation
by adding index $i$ which numbers the spaces in the tensor product:
$$
D_if({\bf g})=\sum_{a,b}e_{ab}^{(i)}
\frac{\p}{\p \varepsilon}f\left (e^{\varepsilon {\bf e}_{ba}}{\bf g}
\right )\Bigr |_{\varepsilon =0}.
$$
Here $e_{ab}^{(i)}$ acts non-trivially in $V_i$. 
In this notation we have, for example: 
$D_1 \mbox{tr}\, {\bf g}={\bf g}_1$, $D_2 {\bf g}_1={\sf P}_{21}{\bf g}_1$, while the relation
(\ref{co2}) is written in the form $[D_2, D_1]={\sf P}_{12}(D_1-D_2)$.

A careful analysis shows that the transfer matrix
${\sf T}_{\lambda}(u)$ can be expressed as
\beq\label{co3}
{\sf T}_{\lambda}(x)=(x-x_N+\eta D_N)\ldots (x-x_1+\eta D_1)\chi_{\lambda}({\bf g}).
\eeq
With the help of this representation, the authors of
\cite{KV08} managed to prove 
the CBR identities 
for models with rational $R$-matrices 
in a direct way.
Here we only mention that this representation allows one to obtain 
the expression for the quantum determinant in a very easy way:
in this case $\chi_{(1^n)}({\bf g})=\det {\bf g}$, and we obtain
\beq\label{co4}
{\sf T}^n (x)=\det {\bf g}
\prod_{j=1}^N(x-x_j+\eta )\cdot {\sf I},
\eeq
where we have used (\ref{co1a}).

\section{The master $T$-operator as an operator-valued
tau-function}

\label{section:master}

\subsection{The master $T$-operator}

It turns out to be instructive to consider 
a generating function for the transfer matrices
${\sf T}_{\lambda}(x)$. As such, it was introduced in \cite{AKLTZ13}
under the name of master $T$-operator (see also \cite{KLT12} for its
preliminary version). 
Let ${\bf t}=\{t_1, t_2, t_3, \ldots \}$ be an infinite set 
of complex variables.  
The master $T$-operator is defined as an infinite series of the form
\beq\label{master10}
{\sf T}(x; {\bf t})=\sum_{\lambda}s_{\lambda}({\bf t}){\sf T}_{\lambda}(x),
\eeq
where the sum, like in (\ref{master5}), is taken over all 
Young diagrams including the empty one. 
Since ${\sf T}_{\lambda}(x)=0$ if $\lambda_1'>n$, the sum
in (\ref{master10}) goes over the diagrams with $\lambda_1'\leq n$.
Like ${\sf T}_{\lambda}(x)$, ${\sf T}(x; {\bf t})$ 
is an operator in the space
${\cal V}$.
It depends on the elements $g_i$ of the (diagonal) 
twist matrix ${\bf g}$ as on parameters.
From commutativity of the ${\sf T}_{\lambda}(x)$ for all
$x, \lambda$ it follows that the operators ${\sf T}(x; {\bf t})$ 
commute for all $x$, ${\bf t}$.
In terms of the co-derivatives, the master
$T$-operator can be represented in the form
\beq\label{master10b}
{\sf T}(x; {\bf t})=(x-x_N+\eta D_N)\ldots (x-x_1+\eta D_1)
\exp \Bigl (\sum_{k\geq 1}t_k {\rm tr}\, {\bf g}^k \Bigr )
\eeq
(to see this, 
one should use (\ref{co3}) and the Cauchy-Littlewood identity (\ref{master5})).

Obviously, ${\sf T}(x; 0)={\sf T}_{\emptyset}(x)=\phi (x)$.
Acting to ${\sf T}(x; {\bf t})$ 
by differential operators in $t_k$ at ${\bf t}=0$,
one can reproduce all the transfer matrices
${\sf T}_{\lambda}(x)$. For example,
\beq\label{master10a}
{\sf T}_{(1)}(x)=\p_{t_1}{\sf T}(x; {\bf t})\Bigr |_{{\bf t}=0}, \qquad
{\sf T}_{(2)}(x)=\frac{1}{2}\left ( \p_{t_1}^2 +\p_{t_2}\right ){\sf T}(x; {\bf t})\Bigr |_{{\bf t}=0}.
\eeq
The general formula is a direct consequence of orthogonality of the
Schur functions which can be easily 
derived from the Cauchy-Littlewood identity in the form
$$
s_{\lambda}(\tilde \p )s_{\mu}({\bf t})
\Bigr |_{{\bf t}=0}=\delta_{\lambda \mu},
$$
where
$\tilde \p = \{ \p_{t_1}, 
\frac{1}{2}\p_{t_2}, \frac{1}{3}\p_{t_3},\ldots \}$.
Acting by $s_{\lambda}(\tilde \p )$ to the both sides of 
(\ref{master10}), we conclude that
\beq\label{master10c}
{\sf T}_{\lambda}(x)=s_{\lambda}(\tilde \p )
{\sf T}(x; {\bf t})\Bigr |_{{\bf t}=0}.
\eeq

Below we use the standard notation
\beq\label{notation}
\begin{array}{c}
{\bf t}\pm [z^{-1}]=\Bigl \{ t_1 \pm z^{-1}, t_2 \pm \frac{1}{2}
z^{-2}, t_3 \pm \frac{1}{3} z^{-3}, \ldots \Bigr \}.
\end{array}
\eeq
From (\ref{master10b}) it follows that ${\sf T}(x, 0\pm [z^{-1}])$
is the generating series for the transfer matrices corresponding to
the diagrams of the form of one row or one column:
\beq\label{g1}
{\sf T}(x, [z^{-1}])=\sum_{s=0}^{\infty}z^{-s}{\sf T}_s(x), 
\qquad
{\sf T}(x, -[z^{-1}])=\sum_{a=0}^n(-z)^{-a}{\sf T}^a(x).
\eeq
Indeed, we have for the first equation:
$$
\sum_{s\geq 0}z^{-s}{\sf T}_s(x)=\sum_{s\geq 0}z^{-s}
h_s (\tilde \p ){\sf T}(x; {\bf t})\Bigr |_{{\bf t}=0}=
\exp \Bigl (\sum_{k\geq 1}\frac{1}{k}z^{-k}\p_{t_k}\Bigr )
{\sf T}(x; {\bf t})\Bigr |_{{\bf t}=0}=
{\sf T}(x, [z^{-1}]),
$$
where we have used the generating function (\ref{he1}). 
The second identity is proved in a similar way.

More generally, from (\ref{master10b}) we also have:
\beq\label{master10d}
\begin{array}{l}
\displaystyle{
{\sf T}(x; {\bf t}-[z^{-1}])=
(x\! - \!x_N\! +\! \eta D_N)\ldots (x\! -\! x_1\! +\! \eta D_1)
\Bigl [\det ({\sf I}-z^{-1}{\sf g})
\exp \Bigl (\sum_{k\geq 1}t_k {\rm tr}\, {\bf g}^k \Bigr )\Bigr ],}
\\ \\
\displaystyle{
{\sf T}(x; {\bf t}+[z^{-1}])=
(x\! -\! x_N\! +\! \eta D_N)\ldots (x\! -\! x_1\! +\! \eta D_1)
\Bigl [\frac{1}{\det ({\sf I}-z^{-1}{\sf g})}
\exp \Bigl (\sum_{k\geq 1}t_k {\rm tr}\, {\bf g}^k \Bigr )\Bigr ].}
\end{array}
\eeq
From the right-hand side of the second formula it is clear
that ${\sf T}(x; {\bf t}+[z^{-1}])$ has poles 
if $z$ equals any eigenvalue of ${\bf g}$. Since each co-derivative
raises the order of a pole, these poles are multiple.

\subsection{Bilinear functional relations for the master
$T$-operator}

Let us pass to the most important property of the master 
$T$-operator, which establishes a close connection with the theory of classical
integrable nonlinear partial differential equations.
Namely, it was proved in \cite{AKLTZ13} that the CBR relations
(\ref{master8a}) (or (\ref{master9a})) mean that any eigenvalue 
of the master $T$-operator as a function of the variables
$x$, ${\bf t}$ is a tau-function of the classical 
modified KP (mKP) hierarchy known in the theory of soliton equations. 
This follows from the fact, proved in \cite{AKLTZ13}
(see also \cite{AZ13}), that if the 
coefficients $c_{\lambda}(x)$ 
in the expansion
\beq\label{g2}
\tau (x, {\bf t})=\sum_{\lambda} c_{\lambda}(x) s_{\lambda}({\bf t})
\eeq
obey the quantum Jacobi-Trudi relations
\beq\label{master8c}
c_{\lambda}(x)=
\left (\prod_{k=1}^{\lambda_1'-1}c_{\emptyset}(x-k\eta )\right )^{-1}
\det_{1\leq i,j\leq \lambda_1'}c_{(\lambda_i-i+j)}
(x-(j\! -\! 1)\eta ),
\eeq
\beq\label{master9c}
c_{\lambda}(x)=
\left (\prod_{k=1}^{\lambda_1-1}c_{\emptyset} (x+k\eta )
\right )^{-1}\det_{1\leq i,j
\leq \lambda_1}c_{(1^{\lambda_i'-i+j})}(x+(j\! -\! 1)\eta ),
\eeq
then $\tau (x, {\bf t})$
is a tau-function of the mKP hierarchy, in which $x$ is identified 
with the ``zeroth time''. This fact allows one to regard the
master $T$-operator (\ref{master10}) as an operator-valued tau-function
meaning that any its eigenvalue is a solution to the mKP hierarchy
represented in the bilinear form.

Note that there is a freedom to
multiply the tau-function by $A^x$, where $A$ is any
non-zero constant:
the transformation $c_{\lambda}(x)\to A^x c_{\lambda}(x)$ 
does not spoil 
the relations (\ref{master8c}), (\ref{master9c}).
We call the tau-functions that differ by a factor of the form
$A^x$ equivalent. 

It follows from the bilinear formalism \cite{JM83,DJKM83} that
the generating bilinear relation for the master $T$-operator
can be written in the form
\beq\label{master11}
\begin{array}{l}
\displaystyle{
\oint_{C_{\infty}}z^{(x-x')/\eta}
e^{\xi ({\bf t}-{\bf t}', z)}
{\sf T}\left (x; {\bf t}-[z^{-1}]\right ){\sf T}\left (x' ; 
{\bf t}'+[z^{-1}]\right )dz}
\\ \\
\phantom{aaaaaaaaaaaaaaaaaaaaaa}\displaystyle{ =
\delta_{(x-x')/\eta , -1}
{\sf T}(x+\eta ;{\bf t})
{\sf T}(x'-\eta ;{\bf t}'),}
\end{array}
\eeq
which is valid for all ${\bf t}, {\bf t}'$
and $x, x'$ such that $(x-x')/\eta \in \ZZ$ and
$(x-x')/\eta \geq -1$. 
In (\ref{master11}) and in what follows we use the notation
\beq\label{xi}
\xi ({\bf t}, z)=\sum_{k\geq 1}t_kz^k.
\eeq
The integration contour  
$C_{\infty}$ is a big circle of radius $R\to \infty$ 
which separates the singularities coming from
the ${\sf T}$-multipliers and the exponential function. 
For some special values of $x-x'$ and ${\bf t}-{\bf t}'$, 
the integral can be calculated by residue calculus and 
equation (\ref{master11}) is thus 
a source of various bilinear relations
of the Hirota-Miwa type.
For example, one can put $x'=x-\eta$, 
${\bf t}'={\bf t}-[z_1^{-1}]-[z_2^{-1}]$, then
$$
e^{\xi ({\bf t}-{\bf t}',z)}=\frac{z_1z_2}{(z_1-z)(z_2-z)}
$$
and the residue calculus gives 
the following 3-term bilinear 
equation: 
\beq\label{master12}
\begin{array}{c}
z_2{\sf T}\left (x+\eta ; {\bf t}-[z_2^{-1}]\right )
{\sf T}\left (x ; {\bf t}-[z_1^{-1}]\right )-
z_1{\sf T}\left (x+\eta ; {\bf t}-[z_1^{-1}]\right )
{\sf T}\left (x ; {\bf t}-[z_2^{-1}]\right )
\\ \\
+\, (z_1-z_2){\sf T}(x+\eta ; {\bf t}){\sf T}\left (x; {\bf t}-[z_1^{-1}]-[z_2^{-1}]\right )=0.
\end{array}
\eeq
In another form, this equation for 
${\sf T}(x;{\bf t})$ first appeared in \cite{KLT12}.
The same equation arises from (\ref{master11}) 
if one puts $x'=x+\eta$, 
${\bf t}'={\bf t}-[z_1^{-1}]-[z_2^{-1}]$.

We have seen that the behavior of the master 
$T$-operator ${\sf T}(x; {\bf t})$ as a function of ${\bf t}$ in an
infinitesimally small neighborhood of the point 
${\bf t}=0$ contains all information about the transfer matrices
${\sf T}_{\lambda}(x)$.
Below we will see that 
analyzing its analytic properties near 
some other points other than 
${\bf t}=0$, one can recover 
Baxter's $Q$-operators. To be more precise, it was shown in
\cite{AKLTZ13} that
${\sf T}(x; [z_1^{-1}]+\ldots + [z_m^{-1}])$ regarded as a function 
of $z_i$'s has multiple poles when any of the $z_i$'s coincides 
with any one of the twist parameters $g_j$'s and residues at these
poles can be identified with the $Q$-operators. 

\section{The mKP hierarchy}

The mKP hierarchy can be defined as a system of evolution equations
for the Lax operator ${\bf L}(x)$ which is a pseudo-difference 
operator of the form
\beq\label{mkp3}
{\bf L}(x)=e^{\eta \p_x}+\sum_{j\geq 0}u_j(x)e^{-j\eta \p_x}.
\eeq
Here $e^{\pm \eta \p_x}$ is the shift operator acting to functions of
$x$ as
$e^{\pm \eta \p_x}f(x)=f(x\pm \eta )$.
The evolution equations (the Lax equations) in the times $t_k$ 
are as follows:
\beq\label{mkp4}
\p_{t_k}{\bf L}(x) =[{\bf B}_k, \, {\bf L}(x)],
\qquad
{\bf B}_k=({\bf L}^k(x))_{\geq 0},
\eeq
where the notation 
$\displaystyle{\Bigl (\sum_j v_j e^{j\eta \p_x}\Bigr )_{\geq 0}}$ means
$\displaystyle{\sum_{j\geq 0} v_j e^{j\eta \p_x}}$, so
${\bf B}_k$ is a difference operator of order $k$.

An important role in the theory is played by the so-called wave
(or dressing) operator ${\bf W}(x)$ which is a pseudo-difference 
operator of the form
\beq\label{mkp5}
{\bf W}(x)=1+\sum_{j\geq 1}w_j(x)e^{-j\eta \p_x}
\eeq
such that
\beq\label{mkp6}
{\bf L}(x)={\bf W}(x)e^{\eta \p_x}{\bf W}^{-1}(x).
\eeq
This representation is often interpreted as ``dressing'' of the 
trivial Lax operator $e^{\eta \p_x}$ by ${\bf W}$. The wave operator
obeys the evolution equation
\beq\label{mkp7}
\p_{t_k}{\bf W}(x)={\bf B}_k(x){\bf W}(x)-{\bf W}(x)e^{k\eta \p_x},
\eeq
so it depends also on ${\bf t}$: ${\bf W}(x)={\bf W}(x, {\bf t})$.
The inverse operator ${\bf W}^{-1}$ is of the form
\beq\label{mkp9}
{\bf W}^{-1}(x)=1+\sum_{j\geq 1}e^{-j\eta \p_x}w_j^*(x+\eta )=
1+\sum_{j\geq 1}w_j^*(x-(j-1)\eta )e^{-j\eta \p_x}.
\eeq

With the help of the wave operator one can introduce the wave 
function $\psi (x, {\bf t}; z)$, and the adjoint wave function  
$\psi^* (x, {\bf t}; z)$, where $z\in \CC$ is a spectral
parameter:
\beq\label{mkp8}
\begin{array}{l}
\psi (x, {\bf t}; z)={\bf W}(x)z^{x/\eta}e^{\xi ({\bf t}, z)},
\\ \\
\psi^* (x, {\bf t}; z)=({\bf W}^{-1}(x-\eta ))^{\dag}
z^{-x/\eta}e^{-\xi ({\bf t}, z)}.
\end{array}
\eeq
The operation $(\ldots )^{\dag}$ is defined as 
$(f(x)e^{k\eta \p_x})^{\dag}=e^{-k\eta \p_x}f(x)$ and is extended 
by linearity to all pseudo-difference operators. From (\ref{mkp8})
we see that the wave function and its adjoint have the following
expansions as $z\to \infty$:
\beq\label{mkp10}
\begin{array}{l}
\displaystyle{
\psi (x, {\bf t}; z)=z^{x/\eta}e^{\xi ({\bf t}, z)}
\Bigl (1+ \frac{w_1(x, {\bf t})}{z}+\frac{w_2(x, {\bf t})}{z^2}+
\ldots \Bigr ),}
\end{array}
\eeq
\beq\label{mkp10a}
\begin{array}{l}
\displaystyle{
\psi^* (x, {\bf t}; z)=z^{-x/\eta}e^{-\xi ({\bf t}, z)}
\Bigl (1+ \frac{w_1^*(x,{\bf t})}{z}+
\frac{w_2^*(x,{\bf t})}{z^2}+\ldots \Bigr ).}
\end{array}
\eeq

It can be proved that there exists a function $\tau (x, {\bf t})$
such that the wave function and its adjoint are expressed through
it in the following way:
\beq\label{BA1}
\psi (x, {\bf t};z)=z^{x/\eta}e^{\xi ({\bf t}, z)}
\frac{\tau (x, {\bf t}-[z^{-1}])}{\tau (x , {\bf t})},
\eeq
\beq\label{BA2}
\psi^* (x, {\bf t};z)=z^{-x/\eta}e^{-\xi ({\bf t}, z)}
\frac{\tau (x, {\bf t}+[z^{-1}])}{\tau (x , {\bf t})},
\eeq
where we use the notation introduced in (\ref{notation}).
This function is called the {\it tau-function}. It plays a 
fundamental role in the theory because equations (\ref{BA1}),
(\ref{BA2}) allow one to express all the coefficients 
$w_j(x, {\bf t})$ of the wave operator
(and thus the coefficients $u_j(x, {\bf t})$ of the Lax operator)
in terms of it.

The tau-function satisfies the integral bilinear equation 
of the form (\ref{master11}):
\beq\label{master11a}
\begin{array}{l}
\displaystyle{
\oint_{C_{\infty}}z^{(x-x')/\eta}
e^{\xi ({\bf t}-{\bf t}', z)}
\tau \left (x; {\bf t}-[z^{-1}]\right )\tau \left (x' ; 
{\bf t}'+[z^{-1}]\right )dz}
\\ \\
\phantom{aaaaaaaaaaaaaaaaaaaaaa}\displaystyle{ =
\delta_{(x-x')/\eta , -1}
\tau (x+\eta ;{\bf t})
\tau (x'-\eta ;{\bf t}')}
\end{array}
\eeq
valid for all ${\bf t}, {\bf t}'$ and $x,x'$ such that
$(x-x')/\eta \in \ZZ_{\geq -1}$.
Its corollary is the equation of the Hirota-Miwa type:
\beq\label{master12a}
\begin{array}{c}
z_2\tau \left (x+\eta ; {\bf t}-[z_2^{-1}]\right )
\tau \left (x ; {\bf t}-[z_1^{-1}]\right )-
z_1\tau \left (x+\eta ; {\bf t}-[z_1^{-1}]\right )
\tau \left (x ; {\bf t}-[z_2^{-1}]\right )
\\ \\
+\, (z_1-z_2)\tau \Bigl (x+\eta ; {\bf t}\Bigr )\tau 
\left (x; {\bf t}-[z_1^{-1}]-[z_2^{-1}]\right )=0.
\end{array}
\eeq

The wave function and its adjoint satisfy an infinite
number of differential-difference equations. The simplest ones are
\beq\label{diff1}
\begin{array}{l}
\phantom{-}\p_{t_1}\psi (x, {\bf t};z)=\psi (x+\eta ,{\bf t};z)+
v(x, {\bf t})\psi (x, {\bf t};z), 
\\ \\
-\p_{t_1}\psi^* (x, {\bf t};z)=\psi^* (x-\eta ,{\bf t};z)+
v(x-\eta , {\bf t})\psi^* (x, {\bf t};z),
\end{array}
\eeq
where
\beq\label{diff2}
v(x, {\bf t})=\p_{t_1}\log 
\frac{\tau (x+\eta ,{\bf t} )}{\tau (x, {\bf t})}.
\eeq
After the substitutions (\ref{BA1}), (\ref{BA2})  
both of them become equivalent to the following bilinear
equation for the tau-function:
\beq\label{diff3}
\begin{array}{l}
\phantom{a}z\tau \Bigl (x+\eta , {\bf t}\Bigr )
\tau \Bigl (x , {\bf t}-[z^{-1}]\Bigr )-
z\tau \Bigl (x, {\bf t}\Bigr )
\tau \Bigl (x+\eta , {\bf t}-[z^{-1}]\Bigr )
\\ \\
=\tau \Bigl (x , {\bf t}-[z^{-1}]\Bigr )
\p_{t_1}\tau \Bigl (x+\eta , {\bf t}\Bigr )
-\tau \Bigl (x+\eta , {\bf t}\Bigr )
\p_{t_1}\tau \Bigl (x , {\bf t}-[z^{-1}]\Bigr ),
\end{array}
\eeq
which follows from (\ref{master12a}) in the limit $z_2\to \infty$
if one puts $z_1=z$. 

The mKP hierarchy has a lot of solutions of very different nature.
To specify the class of solutions to which eigenvalues 
of transfer matrices
of quantum spin chains belong, 
it is important to note that for $GL(n)$-invariant 
models the sum in the series (\ref{g2}) goes over Young diagrams
$\lambda$ with not more than $n$ non-empty rows:
\beq\label{g2a}
\tau (x, {\bf t})=\sum_{\lambda , \,
\lambda_1'\leq n} c_{\lambda}(x) s_{\lambda}({\bf t}).
\eeq
It is easy to see that in this case the series in inverse powers
of $z$ for $\tau (x, {\bf t}-[z^{-1}])$ (and thus the series
(\ref{mkp10}) for the wave function) truncates at the $n$-th term.
(This is also obvious from (\ref{master10d}).)
As far as analytical properties in $z$ of the function 
$\tau (x, {\bf t}+[z^{-1}])$ are concerned, the second equation in
(\ref{master10d}) suggests that it has multiple poles when
$z$ is equal to any eigenvalue of the twist matrix.
So, characterizing the relevant class of solutions, 
we should 
take into account that:
\begin{itemize}
\item[a)] The series (\ref{mkp10}) for the wave function 
truncates at the $n$-th term (i.e., the wave function, as a function
of $z$, has pole of order $n$ at $z=0$);
\item[b)] The tau-function is a polynomial in $x$ of degree $N$.
\item[c)] The adjoint wave function has multiple poles at some
points $p_i \in \CC$, $i=1, \ldots , n$.
\end{itemize}
In the next section we study solutions of this class in more details.

\section{Polynomial solutions to the mKP hierarchy}

\label{section:polynomial}

As it was argued in the previous section,
we are interested in solutions to the mKP hierarchy such that
the tau-function $\tau (x, {\bf t})$ is a polynomial 
or quasi-polynomial\footnote{By quasi-polynomial in $x$ we mean a
polynomial multiplied by $A^x$, where $A$ is a constant.} in $x$.
For the KP equation, such solutions were constructed by Krichever. 
Here we briefly recall this construction, modifying
it for the mKP case.

\subsection{The wave function}

First of all, we consider the wave functions $\psi (x, {\bf t}; z)$
such that the series (\ref{mkp10}) 
in inverse powers of $z$ 
truncates at the $n$-th term:
\beq\label{mkp1}
\psi (x, {\bf t}; z)=z^{x/\eta }e^{\xi ({\bf t}, z)}
\Bigl (1+ \frac{w_1(x, {\bf t})}{z}+\ldots +\frac{w_n(x, {\bf t})}{z^n}
\Bigr ).
\eeq
Let $p_1, \ldots , p_n$ be $n$ distinct points in $\CC$. 
(Later they will be identified with eigenvalues $g_i$ of the twist
matrix ${\bf g}$; the exact relation is $p_i =g_{n-i+1}$,
$i=1, \ldots, n$.)
With each point $p_i$ we associate an integer number 
$M_i\geq 0$ and $n\times (M_i+1)$ rectangular matrix with matrix
elements $a_{i m}$ ($i=1, \ldots , n$, $m=0,1, \ldots , M_i$)
which are parameters of the solution. We assume that $a_{i0}\neq 0$.
With these data at hand, we impose the following $n$ conditions
to the wave function $\psi (x, {\bf t};z)$:
\begin{equation}\label{rat2a}
\sum_{m=0}^{M_i}a_{im}\,
\p_{z}^{m} \psi (x, {\bf t};z)\Bigr |_{z=p_i}\!\! =0\,,
\quad i=1, \ldots , n
\end{equation} 
We call the set of parameters $\{p_i\}$ and $a_{im}$ the Krichever
data of the solution and the conditions (\ref{rat2a}) the 
Krichever conditions. 
Note that the parameters $a_{i0}$,
if they are non-zero,
can always be put equal to 1 without any loss of
generality, so in what follows we assume that $a_{i0}=1$.

The set of conditions (\ref{rat2a})
yields a system of $n$ linear equations for $n$ coefficients
$w_k$ which allows one to fix them as functions of $x, {\bf t}$.
From the general theory of the mKP hierarchy it then follows that
the tau-function associated with
the wave function 
solves the mKP hierarchy.
The coefficients $w_k$ turn out to be rational functions
of their arguments while the tau-function is a polynomial
multiplied by the exponential function of a linear combination
of the variables $x, {\bf t}$ (a quasi-polynomial).
From the algebro-geometric point of view this solution is associated
with a highly singular algebraic curve which is the Riemann sphere
with cusp singularities at the points $p_i$.

It is easy to see that conditions (\ref{rat2a}) are equivalent to
the system of linear equations
\begin{equation}\label{rat3}
A_{i}(x,{\bf t})+\sum_{k=1}^{n}A_{i}(x-k\eta ,{\bf t})w_k=0
\end{equation}
for the coefficients $w_k$ of the wave function, where
\begin{equation}\label{A}
\left. A_{i}(x,{\bf t})=\sum_{m=0}^{M_i}a_{im}
\p_{z}^{m}
\left (z^{x/\eta}e^{\xi ({\bf t},z)}\right ) \right |_{z=p_i}.
\end{equation}
As is seen from this representation,
each function $A_i(x, {\bf t})$ is a polynomial in $x$ of degree
$M_i$ multiplied by the exponential factor $p_i^{x/\eta}$.
Note that $A_i(0,0)=a_{i0}$.
The system (\ref{rat3}) can be solved by applying the
Cramer's rule. This results in the following determinant
representation for the wave function:
\begin{equation}\label{rat5}
\psi (x,{\bf t};z)=z^{x/\eta} e^{\xi ({\bf t},z )}\,
\frac{\det \left (\begin{array}{cccc}
1&z^{-1}&\ldots & z^{-n}\\
A_{1}(x,{\bf t})& A_{1}(x\! -\! \eta ,{\bf t})& \ldots &
A_{1}(x\! -\! n\eta ,{\bf t})\\
\vdots & \vdots &\ddots & \vdots \\
A_{n}(x,{\bf t})&A_{n}(x\! -\! \eta ,{\bf t})& \ldots
&A_{n}(x\! -\! n\eta ,{\bf t})
\end{array}\right )}{\det \left (\begin{array}{ccc}
A_{1}(x-\eta ,{\bf t})&\ldots & A_{1}(x-n\eta ,{\bf t})\\
\vdots &\ddots & \vdots \\
A_{n}(x-\eta ,{\bf t})& \ldots & A_{n}(x-n\eta ,{\bf t})
\end{array}\right )}.
\end{equation}
From the definition (\ref{A}) we have;
\begin{equation}\label{mkp2}
\begin{array}{ll}
    A_{k}(x,{\bf t}-[z^{-1}])=&
    \displaystyle{\left. \sum_{m=0}^{M_k}a_{km}
      \p_{\zeta}^{m} \left (\zeta^{x/\eta}e^{\xi ({\bf t},\zeta )}
      \Bigl (1-\frac{\zeta}{z}\Bigr ) \right ) \right |_{\zeta =p_i}}
        \\ & \\
        & =\, 
    A_k(x,{\bf t})-A_k(x+\eta , {\bf t})z^{-1}.
  \end{array}
\end{equation}
Using this, it is straightforward to verify
that equation (\ref{rat5}) agrees with the general relation (\ref{BA1}),
with the tau-function being
given by the determinant in the denominator:
\begin{equation}\label{rat6}
\tau (x,{\bf t})=\det_{i,j=1, \ldots , n}A_{i}(x\! -\! j\eta,\, {\bf t})
=\det \left (\begin{array}{ccc}
A_{1}(x-\eta ,{\bf t})&\ldots & A_{1}(x-n\eta ,{\bf t})\\
\vdots &\ddots & \vdots \\
A_{n}(x-\eta ,{\bf t})& \ldots & A_{n}(x-n\eta ,{\bf t})
\end{array}\right ).
\end{equation}
It is a polynomial in $x$ of degree
$N=\sum\limits_{j=1}^{n}M_j$ multiplied by
$\prod\limits_{i=1}^{n}p_{i}^{x/\eta}e^{\xi ({\bf t}, p_i)}$.

It follows from (\ref{rat5}) that the last coefficient
in (\ref{mkp1}), $w_n$,
is given by
\begin{equation}\label{rat7}
w_{n}(x,{\bf t})=(-1)^N
\frac{\tau (x+\eta ,{\bf t})}{\tau (x, {\bf t})}.
\end{equation}
We also note the property
\begin{equation}\label{rat4b}
\p_{t_1}A_{i}(x,{\bf t})=A_{i}(x+\eta ,{\bf t})
\end{equation}
from which one can see that the first coefficient
in (\ref{mkp1}), $w_1$,
is given by
\begin{equation}\label{rat7a}
w_{1}(x,{\bf t})=-\p_{t_1} \log \tau (x, {\bf t}).
\end{equation}

\subsection{The adjoint wave function}

Similarly to (\ref{mkp2}), we have:
\begin{equation}\label{rat8}
A_{k}(x,{\bf t}+[z^{-1}])=
\left. \sum_{m=0}^{M_k}a_{km}
\p_{\zeta}^{m}
\left (\frac{\zeta^{x/\eta}e^{\xi ({\bf t},\zeta )}}{1-\zeta/z}
\right ) \right |_{\zeta =p_k},
\end{equation}
Note that this function regarded as a function of $z$
has a pole of order $M_k+1$ at $z=p_k$. The principal term is
\begin{equation}\label{rat9}
A_{k}(x,{\bf t}+[z^{-1}])=
\frac{M_k! \, a_{kM_k}p_{k}^{x/\eta +1}
e^{\xi ({\bf t}, p_k)}}{(z-p_k)^{M_k+1}}\, +
\, \ldots \, .
\end{equation}
We also see from (\ref{rat8}) that 
the function
$A_{k}(x,{\bf t}+[z^{-1}])$ is a rational function of $z$ 
with simple zero at $z=0$.

In order to obtain a more detailed information about the pole structure
of this function, let us add and subtract the term
$z^{x/\eta} e^{\xi ({\bf t},z)}$
in the numerator in (\ref{rat8}) 
and separate the nonsingular part 
from the singular one:
\beq\label{rat8b}
\begin{array}{l}
\displaystyle{
A_{k}(x,{\bf t}+[z^{-1}])= z^{x/\eta +1}
e^{\xi ({\bf t},z)} \sum_{m=0}^{M_k}
\frac{m! \, a_{km}}{(z-p_k )^{m+1}}}
\\ \\
\phantom{aaaaaaaaaaaaaaaaaaaa}
\displaystyle{
\left.-
z \Bigl (\sum_{m=0}^{M_k} a_{km}\p_{\zeta}^{m} \Bigr )\,
\frac{z^{x/\eta} e^{\xi ({\bf t},z)}\! -\!
\zeta^{x/\eta} e^{\xi ({\bf t},\zeta )}}{z-\zeta}\, 
\right |_{\zeta =p_k}.}
\end{array}
\eeq
The first sum gives the multiple pole structure at the point \(p_k\)
while the second term
is obviously regular at $p_k$ and has possible (essential)
singularities and branching only at $0$ and $\infty$.

Rewriting (\ref{mkp2}) in the form
$$
A_k(x, {\bf t}+[z^{-1}])=A_k(x,{\bf t})+
z^{-1}A_k(x+\eta , {\bf t}+[z^{-1}]),
$$
it is straightforward to check that
\begin{equation}\label{rat10}
\tau (x, {\bf t}+[z^{-1}])=\det \left (\begin{array}{cccc}
A_{1}\left (x\! -\! \eta ,{\bf t}\! +\! [z^{-1}]\right )&
A_{1}(x\! -\! 2\eta ,{\bf t})&
\ldots & A_{1}(x\! -\! n\eta ,{\bf t})\\
A_{2}\left (x\! -\! \eta ,{\bf t}\! +\! [z^{-1}]\right )&
A_{2}(x\! -\! 2\eta ,{\bf t})&
\ldots & A_{2}(x\! -\! n\eta ,{\bf t})\\
\vdots & \vdots &\ddots & \vdots \\
A_{n}\left (x\! -\! \eta ,{\bf t}\! +\! [z^{-1}]\right )&
A_{n}(x\! -\! 2\eta ,{\bf t})&
\ldots &A_{n}(x\! -\! n\eta ,{\bf t}).
\end{array}\right ).
\end{equation}
Expanding (\ref{rat8}) in powers of $z$, we get:
$$
A_{k}(x,{\bf t}+[z^{-1}]) =A_k(x,{\bf t})+
A_k(x+\eta ,{\bf t})z^{-1}+A_k(x+2\eta ,{\bf t})z^{-2}+\ldots \, ,
$$
and so the expansion of $\tau (x,{\bf t}+[z^{-1}])$ around
$\infty$ reads
\begin{equation}\label{rat11}
\tau (x, {\bf t}+[z^{-1}])=\sum_{s=0}^{\infty}z^{-s}
\det \left (\begin{array}{cccc}
A_{1}\left (x\! +(s-1)\eta ,{\bf t}\right )&
A_{1}(x\! -\! 2\eta ,{\bf t})&
\ldots & A_{1}(x\! -\! n\eta ,{\bf t})\\
A_{2}\left (x\! +(s-1)\eta ,{\bf t}\right )& A_{2}(x\! -\! 2\eta ,{\bf t})&
\ldots & A_{2}(x\! -\! n\eta ,{\bf t})\\
\vdots & \vdots &\ddots & \vdots \\
A_{n}\left (x\! +(s-1)\eta ,{\bf t}\right )&A_{n}(x\! -\! 2\eta ,{\bf t})&
\ldots &A_{n}(x\! -\! n\eta ,{\bf t})
\end{array}\right ).
\end{equation}
We thus see that the adjoint wave function has the
determinant representation
\begin{equation}\label{rat12}
\psi^{*}(x,{\bf t};z)=z^{-x/\eta}e^{-\xi ({\bf t},z)}\,
\frac{\det \left (\begin{array}{cccc}
A_{1}\left (x\! -\! \eta ,{\bf t}\! +\! [z^{-1}]\right )&
A_{1}(x\! -\! 2\eta ,{\bf t})&
\ldots & A_{1}(x\! -\! n\eta ,{\bf t})\\
A_{2}\left (x\! -\! \eta ,{\bf t}\! +\! [z^{-1}]\right )&
A_{2}(x\! -\! 2\eta ,{\bf t})&
\ldots & A_{2}(x\! -\! n\eta ,{\bf t})\\
\vdots & \vdots &\ddots & \vdots \\
A_{n}\left (x\! -\! \eta ,{\bf t}\! +\! [z^{-1}]\right )&
A_{n}(x\! -\! 2\eta ,{\bf t})&
\ldots &A_{n}(x\! -\! n\eta ,{\bf t})
\end{array}\right )}{\det \left (\begin{array}{cccc}
A_{1}\left (x\! -\! \eta ,{\bf t}\right )& A_{1}(x\! -\! 2\eta ,{\bf t})&
\ldots & A_{1}(x\! -\! n\eta ,{\bf t})\\
A_{2}\left (x\! -\! \eta ,{\bf t}\right )& A_{2}(x\! -\! 2\eta ,{\bf t})&
\ldots & A_{2}(x\! -\! n\eta ,{\bf t})\\
\vdots & \vdots &\ddots & \vdots \\
A_{n}\left (x\! -\! \eta ,{\bf t}\right )&A_{n}(x\! -\! 2\eta ,{\bf t})&
\ldots &A_{n}(x\! -\! n\eta ,{\bf t})
\end{array}\right ).}
\end{equation}
This function has multiple poles at the points $p_i$.
Since the function
$A_{k}(x,{\bf t}+[z^{-1}])$ has a simple zero at $z=0$,
it is clear from
(\ref{rat6}) that
the function $\tau (x,{\bf t} +[z^{-1}])$ and thus the function
$z^{x/\eta} e^{\xi ({\bf t},z)}
\psi^{*}(x, {\bf t};z)$ has zero of order $n$ at $z=0$.

Let us introduce the notation
\begin{equation}\label{rat13a}
\hat A_{k}(x,{\bf t}):= \det_{
{i=1, \,\ldots , \not k , \ldots , N\atop j=1,\, \ldots \, , \, n-1}}
\, A_{i}(x+(1-j)\eta , \, {\bf t})
\end{equation}
for the minor of the 
$n\! \times \! n$ matrix $A_{i}(x+(1-j)\eta )$, $1\leq i,j\leq n$.
Then, expanding the determinant in the numerator of
(\ref{rat12}) in the first column, we obtain:
$$
\psi^{*}(x, {\bf t};z)=\frac{z^{-x/\eta}
e^{-\xi ({\bf t},z)}}{\tau (x,{\bf t})}
\sum_{k=1}^{n} (-1)^{k-1}
\hat A_{k}(x-2\eta , {\bf t}) \,A_k (x-\eta , {\bf t}+[z^{-1}]).
$$
Using (\ref{rat8b}) we can extract the poles:
\begin{equation}\label{rat13b}
\psi^{*}(x, {\bf t};z)=\sum_{k=1}^{n}(-1)^{k-1}
\frac{\hat A_{k}(x-2\eta , {\bf t})}{\tau (x, {\bf t})}
\sum_{m=0}^{M_k}\frac{m! \, a_{km}}{(z-p_k )^{m+1}} \,
+\, \mbox{terms regular at all $p_k$}.
\end{equation}
Therefore,
\begin{equation}\label{rat13c}
\res_{z=p_k}\left [ (z-p_k)^m
\psi^{*}(x, {\bf t};z)\right ]=(-1)^{k-1}m! \, a_{km} \,
\frac{\hat A_{k}(x-2\eta , {\bf t})}{\tau (x, {\bf t})}
\end{equation}
for $m=0, 1, \ldots , M_k$, or, in terms of the tau-function,
\beq\label{rat13d}
\res_{z=p_k}\left [ (z-p_k)^m z^{-x/\eta}e^{-\xi ({\bf t}, z)}
\tau (x, {\bf t}+[z^{-1}])\right ]=(-1)^{k-1}m! \, a_{km} \,
\hat A_{k}(x-2\eta , {\bf t}).
\eeq
Note that the expression in the right-hand side has the same structure
as (\ref{rat6}) and, therefore, is a (quasi)polynomial tau-function.
It is constructed by means of the Krichever 
data from which the point $p_k$ is excluded
in the same way as $\tau (x,{\bf t})$.
Thus the passage from
$\tau (x,{\bf t})$ to $A_{k}(x, {\bf t})$ can be regarded as
a B\"acklund transformation.

\subsection{Undressing transformations}

Our main goal is to show that the nested Bethe ansatz scheme 
is equivalent to a chain of some
special B\"acklund transformations of the initial
polynomial mKP solution with the Krichever 
data $p_1, \ldots , p_n$,
$a_{im}$
that ``undress'' it to the trivial solution by reducing the
number of the points $p_i$ in succession.
Here we present the idea solely in terms of the mKP
hierarchy.

Removing a point from the Krichever data
of a polynomial solution is a B\"acklund transformation.
It sends a (quasi)polynomial tau-function to another one.
Moreover, we are going to consider a chain of such transformations.
To this end, let us fix some order in the set of points 
$p_1, p_2, \ldots , p_n$ and remove first $p_n$, then $p_{n-1}$
and so on, up to $p_1$. (Finally, removing all the points, we obtain
the trivial solution which is a constant.)
The key equation that allows one to implement
such transformations is (\ref{rat13d}). 
Let us consider its $m=0$ case and start from removing the point $p_n$.
Specifically, consider the function
\begin{equation}\label{Back1}
\tau ^{(n-1)}(x,{\bf t})=(-1)^{n-1}\res_{z=p_n}\left (
z^{-x/\eta -1}e^{-\xi ({\bf t},z)}\tau (x+\eta ,{\bf t}+[z^{-1}])
\right ).
\end{equation}
According to the $m=0$ case of (\ref{rat13d}),
it is equal to
\begin{equation}\label{Back2}
\tau ^{(n-1)}(x, {\bf t})=
\det \left (
\begin{array}{cccc}
A_{1}\left (x\! -\! \eta ,{\bf t}\right )& 
A_{1}\, (x\! -\! 2\eta ,{\bf t})&
\ldots & A_{1}(x\! -\! (n-1)\eta ,{\bf t})\\
A_{2}\left (x\! -\! \eta ,{\bf t}\right )& 
A_{2}\, (x\! -\! 2\eta ,{\bf t})&
\ldots & A_{2}(x\! -\! (n-1)\eta ,{\bf t})\\
\vdots & \vdots &\ddots & \vdots \\
A_{n-1}\left (x\! -\! \eta ,{\bf t}\right )&A_{n-1}(x\! -\! 
2\eta ,{\bf t})&
\ldots &A_{n-1}\, (x\! -\! (n\! -\! 1)\eta ,{\bf t})
\end{array}
\right ),
\end{equation}
i.e., to the minor $M_{n,n}$ of the matrix $A_{i}(x+(1-j)\eta )$.
Therefore, it is 
a tau-function, i.e., it
satisfies the same bilinear
equations of the mKP hierarchy 
as $\tau (x, {\bf t})$ does and
$\tau^{(n)} \rightarrow \tau^{(n-1)}$,
where $\tau^{(n)} (x, {\bf t})=\tau (x, {\bf t})$
is the initial member of the undressing chain, is indeed a B\"acklund
transformation. 

Having at hand $\tau^{(n-1)}(x, {\bf t})$, one can construct the wave
function
\begin{equation}\label{Back3}
\psi^{(n-1)}(x, {\bf t}; z)=z^{x/\eta} e^{\xi ({\bf t},z)}
\, \frac{\tau^{(n-1)}(x, {\bf t} -[z^{-1}])}{\tau^{(n-1)}(x, {\bf t})}
\end{equation}
which obeys the same Krichever conditions (\ref{rat2a})
at the points $p_1, \ldots ,p_{n-1}$ but not at the point
$p_n$,
where no condition is imposed. Note that the wave function
$\psi^{(n-1)}$ has the form similar to (\ref{mkp1}) but with a pole 
at $z=0$ of order $n-1$ rather than $n$, so the number
of conditions again matches the number of unknown coefficients.
The adjoint wave function is
\begin{equation}\label{Back4}
\psi^{*(n-1)}(x, {\bf t}; z)=z^{-x/\eta} e^{-\xi ({\bf t},z)}
\, \frac{\tau^{(n-1)}(x, {\bf t} +[z^{-1}])}{\tau^{(n-1)}(x, {\bf t})}.
\end{equation}
It has multiple poles at the points
$p_1, \ldots ,p_{n-1}$.

This process can be continued by taking
the residue of $z^{-x/\eta -1}e^{-\xi ({\bf t},z)}
\tau^{(n-1)}(x+\eta ,{\bf t} +[z^{-1}])$
at the next point $p_{n-1}$ and introducing the function
\beq\label{Back5}
\begin{array}{c}
\tau^{(n-2)}(x, {\bf t})=
(-1)^{n-2}
\res_{z=p_{n-1}}\left (
z^{-x/\eta -1}e^{-\xi ({\bf t},z)}
\tau^{(n-1)}(x+\eta ,{\bf t}+[z^{-1}])
\right )
\\ \\
\displaystyle{
=\res_{z_{n-1}=p_{n-1}\atop z_{n}=p_{n}}
\Bigl ((z_{n-1}z_n)^{-x/\eta -2} e^{-\xi ({\bf t}, z_{n-1})-
\xi ({\bf t}, z_{n})}(z_n-z_{n-1}) 
\tau^{(n)}(x+2\eta , {\bf t}+[z_{n-1}^{-1}]+[z_{n}^{-1}])\Bigr )}
\\ \\
=\, \det\limits_{r,s = 1, \ldots , n-2}
\Bigl [A_r (x-s\eta, {\bf t})\Bigr ]
\end{array}
\eeq

We thus obtain a chain of B\"acklund  
transformations
\beq\label{chain}
\tau =\tau^{(n)} \rightarrow \tau^{(n-1)} \rightarrow \tau^{(n-2)}
\rightarrow \tau^{(n-3)}\rightarrow \ldots \rightarrow
\tau^{(1)}\rightarrow \tau^{(0)}=1
\eeq
which ``undress'' the initial solution up to the trivial one.
The general recursive formula at the $m$-th level is
the same as (\ref{Back1}) with the change $n\to m$.
Solving the recursion relation, we find:
\beq\label{Back5gen}
\begin{array}{l}
\displaystyle{
\tau^{(m)}(x, {\bf t})
=\res_{z_j=p_j\atop j=m+1, \ldots , n}
\left [\prod_{\alpha =m+1}^{n}(z_{\alpha}^{-x/\eta +m-n}
e^{\xi ({\bf t}, z_{\alpha})})
\Delta (z_{m+1}, \ldots , z_{n})\right.}
\\ \\
\phantom{aaaaaaaaaaaaaaa}\displaystyle{
\left. \times \, \tau \Bigl (x+(n-m)\eta , {\bf t}+\!\!
\sum_{\alpha =m+1}^n [z_{\alpha}^{-1}]\Bigr )\right ],}
\end{array}
\eeq
where
$\displaystyle{
\Delta (x_1, \ldots , x_m)=\prod\limits_{i>j}^m (x_i-x_j)}$
is the Vandermonde determinant.
In particular, on the previous to the last level we have
\beq\label{undress1}
\tau^{(1)}(x, {\bf t})
=A_1(x-\eta , {\bf t}) 
\eeq
and on the last level
$
\tau^{(0)}({\bf t})=1$. The function $\tau^{(m)}(x, {\bf t})$ is
a quasipolynomial in $x$ of degree $N_m=M_1+\ldots +M_{m}$.

Note that equation (\ref{rat13d}) allows one to make 
the same B\"acklund transformations by picking the coefficients
in front of higher order poles of the function 
$\tau (x, {\bf t}+[z^{-1}])$ at $z=p_1, \ldots , p_n$. 
The results differ from (\ref{Back5gen}) 
by normalization factors independent 
of $x, {\bf t}$.

\subsection{The dressing chain}

\label{section:dressing}

In this section we will follow the arrows
of the ``undressing chain'' (\ref{chain})
in the inverse direction, from right to left. Then each step is 
naturally referred to as a ``dressing'' transformation, which 
allows one to construct more complicated  
tau-functions from simpler
ones. 

The wave function $\psi^{(m)}(x, {\bf t}; z)$ at the $m$-th step
of the chain is given by (\ref{Back3}) with the change $n\to m$.
We claim that this wave function can be obtained from the previous
one, $\psi^{(m-1)}(x, {\bf t}; z)$, by action of a first order 
difference operator. Namely, we have:
\beq\label{d1}
\psi^{(m)}(x, {\bf t}; z)=
\left (1-\frac{\tau^{(m)}(x+\eta )\tau^{(m-1)}(x-\eta )}{\tau^{(m)}(x+\eta )
\tau^{(m-1)}(x)}\, e^{-\eta \p_x}\right )
\psi^{(m-1)}(x, {\bf t}; z),
\eeq
where the dependence of the tau-functions 
on ${\bf t}$ is not shown explicitly. To prove this equality, we
write down the right-hand side,
$$
z^{x/\eta}e^{\xi ({\bf t}, z)}\left [
\frac{1}{\tau^{(m-1)}(x)}\,
\det \left (\begin{array}{cccc}
1&z^{-1}&\ldots & z^{-(m-1)}\\
A_{1}(x)& A_{1}(x\! -\! \eta )& \ldots &
A_{1}(x\! -\! (m-1)\eta )\\
\vdots & \vdots &\ddots & \vdots \\
A_{m-1}(x)&A_{m-1}(x\! -\! \eta )& \ldots
&A_{m-1}(u\! -\! n\eta )
\end{array}\right )\right.
$$
$$
\left. -\, \frac{\tau^{(m)}(x+\eta )}{\tau^{(m)}(x)\tau^{(m-1)}(x)}
\det \left (\begin{array}{cccc}
z^{-1}&z^{-2}&\ldots & z^{-m}\\
A_{1}(x-\eta )& A_{1}(x\! -2\! \eta )& \ldots &
A_{1}(x\! -\! m\eta )\\
\vdots & \vdots &\ddots & \vdots \\
A_{m-1}(x-\eta )&A_{m-1}(x\! -2\! \eta )& \ldots
&A_{m-1}(x\! -\! m\eta )
\end{array}\right )\right ],
$$
and extract the coefficient in front of $z^{-k}$.
It is equal to
$$
w^{(m)}_k=(-1)^k\left [ \frac{\tau^{(m-1), k}(x)}{\tau^{(m-1)}(x)}+
\frac{\tau^{(m)}(x+\eta )
\tau^{(m-1), k-1}(x-\eta )}{\tau^{(m)}(x)\tau^{(m-1)}(x)}\right ],
$$
where
$$
\tau^{(m), k}(x)=\det_{i=1, \ldots , m-1\atop j=0, \ldots , \not k ,
\ldots , m-1}\Bigl (A_i (x-j\eta )\Bigr ).
$$
To show that
\beq\label{d2}
w_k^{(m)}=(-1)^k \frac{\tau^{(m),k}(x)}{\tau^{(m)}(x)},
\eeq
as it should be, we need the determinant identity
\beq\label{d3}
D[j_1 j_2]D[j_3 j_4]+D[j_1 j_4]D[j_2 j_3]=D[j_1 j_3]D[j_2 j_4]
\eeq
valid for determinants $D[ij]$ of square matrices obtained from 
any $m\times (m+2)$ rectangular matrix $M_{ab}$ 
by removing its $i$-th and $j$-th
columns. This identity is one of the Pl\"ucker relations.
Let us take the matrix $M_{ab}$ to be $M_{i1}=\delta_{m,i}$,
$M_{ij}=A_i (x-(j-2)\eta )$ with $i=1, \ldots , m$,
$j=2, \ldots , m+2$, and
$j_1=1$, $j_2=2$, $j_3=k+1$, $j_4=m$. With this choice, 
the identity (\ref{d3}) reads
$$
\tau^{(m-1), k}(x)\tau^{(m)}(x)+
\tau^{(m-1), k-1}(x-\eta )\tau^{(m)}(x+\eta )=
\tau^{(m-1)}(x )\tau^{(m), k}(x)
$$
and thus leads to (\ref{d2}). 

Regarding (\ref{d1}) as a recurrence relation
for the wave functions $\psi^{(m)}$, one can factorize
the wave operator
${\bf W}^{(n)}$ by representing it as a product of $n$ first
order difference operators:
\beq\label{d4}
\begin{array}{c}
\displaystyle{
{\bf W}^{(n)}(x)=\sum_{k=0}^n (-1)^k 
\frac{\tau^{(n),k}(x)}{\tau^{(n)}(x)}\, e^{-k\eta \p_x}}
\\ \\
=\,
(1-U_n(x)e^{-\eta \p_x})(1-U_{n-1}(x)e^{-\eta \p_x})\ldots
(1-U_1(x)e^{-\eta \p_x}),
\end{array}
\eeq
where
\beq\label{d5}
U_m(x)=\frac{\tau^{(m)}(x+\eta )
\tau^{(m-1)}(x-\eta )}{\tau^{(m)}(x )
\tau^{(m-1)}(x)}.
\eeq
In particular, 
\beq\label{d5a}
\frac{\tau^{(n),1}(x)}{\tau^{(n)}(x)}=\sum_{m=0}^n U_m(x)=
\sum_{m=0}^n \frac{\tau^{(m)}(x+\eta )
\tau^{(m-1)}(x-\eta )}{\tau^{(m)}(x)\,
\tau^{(m-1)}(x)}.
\eeq

From the form of the last multiplier in right-hand side of 
(\ref{d4}) it is obvious that
$\Psi (x)=\tau^{(1)}(x+\eta , {\bf t}) =A_1(x, {\bf t})$ is a solution
to the difference equation
\beq\label{d6}
{\bf W}^{(n)}(x)\Psi (x) =0,
\eeq
i.e., it holds
\beq\label{d8}
\sum_{k=0}^n (-1)^k 
\frac{\tau^{(n),k}(x)}{\tau^{(n)}(x)}\, 
\tau^{(1)}(x-(k-1)\eta , {\bf t})=0.
\eeq
The representation of equation (\ref{d6}) in the form
\beq\label{d7}
\det \left (\begin{array}{cccc}
\Psi (x)&\Psi (x-\eta )&\ldots & \Psi (x-n\eta )\\
A_{1}(x)& A_{1}(x\! -\! \eta )& \ldots &
A_{1}(x\! -\! n\eta )\\
\vdots & \vdots &\ddots & \vdots \\
A_{n}(x)&A_{n-1}(x\! -\! \eta )& \ldots
&A_{n}(x\! -\! n\eta )
\end{array}\right )=0
\eeq
makes it obvious that the other $n-1$ solutions are
$A_2(x, {\bf t}), \, A_3(x, {\bf t}), \ldots , A_n(x, {\bf t})$.
(However, this is not so easy to see from (\ref{d4}).)

Let us obtain the relation connecting tau-functions 
at two neighboring levels
of the dressing chain. The easiest way to do this is to
substitute the expression for the wave function
$$
\psi^{(m)}(x, {\bf t}; z)=z^{x/\eta} e^{\xi ({\bf t},z)}
\, \frac{\tau^{(m)}(x, {\bf t} -[z^{-1}])}{\tau^{(m)}(x, {\bf t})}
$$
into (\ref{d1}). This gives the following bilinear equation:
\beq\label{d9}
\begin{array}{l}
\tau^{(m-1)}(x, {\bf t}-[z^{-1}])\tau^{(m)}(x, {\bf t})-
z^{-1}\tau^{(m-1)}(x-\eta , {\bf t}-[z^{-1}])
\tau^{(m)}(x+\eta , {\bf t})
\\ \\
\phantom{aaaaaaaaaaaaaaaaaaaaa}
=
\tau^{(m-1)}(x, {\bf t})\tau^{(m)}(x, {\bf t}-[z^{-1}]).
\end{array}
\eeq
It suggests that the variable $m$ can be interpreted as a discrete
``time'' corresponding to the B\"acklund flow. In (\ref{d9}),
$m$ takes the finite set of integer values $1, 2, \ldots , n$.
It is sometimes convenient to extend this set by including the values
$m=0$ and $m=n+1$; in this case one should put 
$\tau^{(-1)}(x, {\bf t})=\tau^{(n+1)}(x, {\bf t})=0$. 

Let us denote $\tau^{(m)}(x, {\bf t}-[z^{-1}]) =\rho^{(m)}(x)$, then
equation (\ref{d9}) acquires the form
\beq\label{d10}
\rho^{(m-1)}(x)\tau^{(m)}(x)-
z^{-1}\rho^{(m-1)}(x-\eta )
\tau^{(m)}(x+\eta )=
\rho^{(m)}(x)\tau^{(m-1)}(x).
\eeq
We recall that $\tau^{(m)}(x)$ are quasipolynomials in $x$ of degree
$N_m$.
Let $v_{\alpha}^{(m)}$, $\alpha =1, \ldots , N_m$ 
be roots of these quasipolynomials,
then we can write  
\beq\label{d11}
\tau^{(m)}(x)=(p_1\ldots p_m)^{x/\eta}
\prod_{\alpha =1}^{N_m}(x-v_{\alpha}^{(m)}).
\eeq
Putting $x$ equal to $v_{\alpha}^{(m)}$, $v_{\alpha}^{(m)}-\eta$,
$v_{\alpha}^{(m-1)}$ in (\ref{d10}), so that only two of the three
terms survive, we get the system of equations
$$
\left \{
\begin{array}{l}
-z^{-1}\rho^{(m-1)}(v_{\alpha}^{(m)}-\eta )
\tau^{(m)}(v_{\alpha}^{(m)}+\eta )=
\rho^{(m)}(v_{\alpha}^{(m)})\tau^{(m-1)}(v_{\alpha}^{(m)}),
\\ \\
\rho^{(m-1)}(v_{\alpha}^{(m)}-\eta )
\tau^{(m)}(v_{\alpha}^{(m)}-\eta )=
\rho^{(m)}(v_{\alpha}^{(m)}-\eta )\tau^{(m-1)}(v_{\alpha}^{(m)}-\eta ),
\\ \\
\rho^{(m-1)}(v_{\alpha}^{(m-1)} )
\tau^{(m)}(v_{\alpha}^{(m-1)})=z^{-1}
\rho^{(m-1)}(v_{\alpha}^{(m-1)}-\eta )
\tau^{(m)}(v_{\alpha}^{(m-1)}+\eta ).
\end{array}
\right.
$$
Dividing the first equation by the second one and using the
third equation with the shift $m \to m+1$, one can exclude $\rho$.
This yields the following equations for zeros of the tau-functions:
\beq\label{d12}
\frac{\tau^{(m+1)}(v_{\alpha}^{(m)}+\eta )
\tau^{(m)}(v_{\alpha}^{(m)}-\eta )
\tau^{(m-1)}(v_{\alpha}^{(m)})}{\tau^{(m+1)}(v_{\alpha}^{(m)})
\tau^{(m)}(v_{\alpha}^{(m)}+\eta )
\tau^{(m-1)}(v_{\alpha}^{(m)}-\eta )}=-1, \quad \alpha =1, \ldots , N_m,
\eeq
or, plugging here (\ref{d11}),
\beq\label{d13}
\prod_{\beta =1}^{N_{m+1}}
\frac{v_{\alpha}^{(m)}-v_{\beta}^{(m+1)}
+\eta }{v_{\alpha}^{(m)}-v_{\beta}^{(m+1)}}\, 
\prod_{\beta =1, \neq \alpha}^{N_{m}}
\frac{v_{\alpha}^{(m)}-v_{\beta}^{(m)}
-\eta }{v_{\alpha}^{(m)}-v_{\beta}^{(m)} +\eta }\,
\prod_{\beta =1}^{N_{m-1}}
\frac{v_{\alpha}^{(m)}-v_{\beta}^{(m-1)}}{v_{\alpha}^{(m)}-
v_{\beta}^{(m-1)}-\eta }=\frac{p_{m}}{p_{m+1}},
\eeq
in which we recognize the system of 
nested Bethe equations\footnote{We postpone
a detailed comparison with the Bethe equations (\ref{eig2})
obtained via nested Bethe ansatz till the next section.}.
At $m=1$ the last product in the left-hand side should be put
equal to 1.

Note that if equations (\ref{d12}) hold, all coefficients
$\tau^{(n),k}(x, {\bf t})$ in (\ref{d4}) 
are (quasi)\-po\-ly\-no\-mi\-als 
in $x$. Equivalently, equations (\ref{d12}) could be derived 
from the requirement that $\tau^{(n),k}(x, {\bf t})$ are 
quasipolynomials (i.e, that possible poles at the roots 
of $\tau^{(m)}(x, {\bf t})$ for $m=1, \ldots , n-1$ cancel).

Remarkably, equations (\ref{d12}) have the same form as 
equations of motion for the integrable time discretization
of the Ruijsenaars-Schneider model suggested in \cite{NRK96}.
In this interpretation, the Bethe roots $v_{\alpha}^{(m)}$ are
coordinates of the particles at the $m$-th step of discrete time.
In \cite{KLWZ97} it was shown that the same equations can be
obtained from dynamics of zeros of polynomial tau-functions 
subject to the fully difference version of the 
Hirota bilinear equation
(also known as a fully discrete KP equation). However, there are
two important differences between equations (\ref{d12}) and the
equations obtained in \cite{NRK96}. First, in the former the
discrete time $m$ takes only a finite number of values $m=1, 
\ldots , n-1$. Second, the number of particles $N_m$ at 
the $m$-th time step depends on $m$. 

\section{Diagonalization of transfer matrices as a chain
of B\"acklund transformations}

In this section we identify the objects which appeared in the
previous section in the context of the mKP hierarchy with the
standard objects of the theory of quantum spin chains and 
rewrite the key formulas of the previous section in the notation
adopted in the theory of spin chains.

We denote eigenvalues of the master $T$-operator ${\sf T}(x, {\bf t})$
by $T(x, {\bf t})$. Similarly, let $T_{\lambda}(x)$ be eigenvalues
of ${\sf T}_{\lambda}(x)$ (in particular, $T^a(x)$ and $T_s(x)$
are eigenvalues of ${\sf T}^a(x)$ and ${\sf T}_s(x)$ respectively).
The eigenvalues $T(x, {\bf t})$
are connected with the quasipolynomial
tau-functions constructed in the previous section as
\beq\label{e1}
T(x, {\bf t})=(\det {\bf g})^{-x/\eta} \, \tau (x, {\bf t})
\eeq
(the factor $(\det {\bf g})^{-x/\eta}$ is necessary to make
the left-hand side polynomial in $x$ rather than quasipolynomial).
Different eigenvalues correspond to different polynomial 
solutions of the mKP hierarchy subject to the necessary conditions.

The standard objects of the theory of quantum spin chains
can be obtained from the construction of the previous section
at ${\bf t}=0$. For example,
$$
T(x, 0)=\phi (x)=(\det {\bf g})^{-x/\eta}\tau (x, 0),
$$
where $\phi (x)$ is the fixed given polynomial (\ref{phi}) whose
roots are inhomogeneity parameters of the spin chain.
We also note that
$$
T^a(x)=(\det {\bf g})^{-x/\eta}\tau^{(n),a}(x, 0)
$$
with $\tau^{(n),a}(x, {\bf t})$ from (\ref{d4}).

The wave operator ${\bf W}(x, 0)$ in the theory of spin 
chains is known under the name of the non-commutative generating
function for $T^a(x)$:
\beq\label{e2}
{\bf W}(x,0)=\sum_{a=0}^{n}(-1)^a \frac{T^a(x)}{\phi (x)}\,
e^{-a\eta \p_x}.
\eeq
From (\ref{g1}) it follows that ${\bf W}^{-1}(x,0)$ is the
non-commutative generating series for $T_s(x)$:
\beq\label{e3}
{\bf W}^{-1}(x,0)=\sum_{s=0}^{\infty} e^{-s\eta \p_x}\,
\frac{T_s(x+\eta )}{\phi (x+\eta )}=
\sum_{s=0}^{\infty}\frac{T_s(x-(s-1)\eta )}{\phi (x-(s-1)\eta )}\,
e^{-s\eta \p_x}.
\eeq

The functions $\tau^{(m)}(x,0)$ from intermediate levels of the
dressing chain can be identified with eigenvalues of the 
Baxter's $Q$-operators which are quasipolynomials with roots
$v_{\alpha}^{(m)}$:
\beq\label{e4}
\tau^{(m)}(x+\eta ,0)=Q_m(x)=(p_1\ldots p_m)^{x/\eta}
\prod_{\alpha =1}^{N_m}
(x-v_{\alpha}^{(m)}), \quad m=1, \ldots , n-1,
\eeq
where $N_m =M_1 +\ldots +M_m$. The last one, $Q_n(x)$, is the
fixed quasipolynomial $$Q_n(x)=
(\det {\bf g})^{x/\eta}\phi (x).$$

The factorization (\ref{d4}) of the wave operator at ${\bf t}=0$ 
in terms of the $Q_m$'s acquires the form
\beq\label{e6}
\begin{array}{c}
\displaystyle{
{\bf W}(x, 0)=\left (1\! -\! \frac{Q_n(x\! +\! 
\eta )Q_{n-1}(x\! -\! \eta )}{Q_n (x)
Q_{n-1}(x)}e^{-\eta \p_x}\right )
\left (1\! -\! \frac{Q_{n-1}(x\! +\! 
\eta )Q_{n-2}(x\! -\! \eta )}{Q_{n-1}(x)
Q_{n-2}(x)}e^{-\eta \p_x}\right )\ldots }
\\ \\
\displaystyle{
\ldots \times 
\left (1-\frac{Q_{2}(x+\eta )Q_{1}(x-\eta )}{Q_{2}(x)
Q_{1}(x)}e^{-\eta \p_x}\right )
\left (1-\frac{Q_{1}(x+\eta )}{Q_{1}(x)}e^{-\eta \p_x}\right ).}
\end{array}
\eeq
The roots of the $Q_m$'s satisfy the nested Bethe equations 
(\ref{d13}) which in fact follow
from (\ref{e6}) after imposing the condition that
the right-hand side is regular when $x$ is equal to any root
of $Q_1(x), \ldots , Q_{n-1}(x)$. 
In terms of the $Q_m$'s the Bethe equations can be written in the form
\beq\label{d12a}
\frac{Q_{m+1}(v_{\alpha}^{(m)}+\eta )
Q_m(v_{\alpha}^{(m)}-\eta )
Q_{m-1}(v_{\alpha}^{(m)})}{Q_{m+1}(v_{\alpha}^{(m)})
Q_m(v_{\alpha}^{(m)}+\eta )
Q_{m-1}(v_{\alpha}^{(m)}-\eta )}=-1, \quad \alpha =1, \ldots , N_m.
\eeq
In the theory of integrable spin chains, equation 
(\ref{e6}) plays a key role: it allows one to express 
$T^a(x)$ through the eigenvalues of the $Q$-operators
(which can be regarded as known quantities 
as soon as their roots are found from the 
Bethe equations). With the $T^a$'s at hand, 
eigenvalues of all other transfer matrices
${\sf T}_{\lambda}(x)$ can then be found with the help of the
CBR determinant formulas (\ref{master9a}).

Factorization of the conjugated operator, 
\beq\label{e2a}
{\bf W}^{\dag}(x,0)=\sum_{a=0}^n (-1)^a 
\frac{T^a(x+a\eta )}{\phi (x+a\eta )}\, e^{a\eta \p_x},
\eeq
immediately follows from (\ref{e6}):
\beq\label{e6a}
\begin{array}{c}
\displaystyle{
{\bf W}^{\dag}(x-\eta , 0)=
\left (1-\frac{Q_{1}(x+\eta )}{Q_{1}(x)}e^{\eta \p_x}\right )
\left (1-\frac{Q_{2}(x+\eta )Q_{1}(x-\eta )}{Q_{2}(x)Q_{1}(x)}
e^{\eta \p_x}\right )\ldots }
\\ \\
\displaystyle{
\ldots \times 
\left (1\! -\! \frac{Q_{n-1}(x\! +\! 
\eta )Q_{n-2}(x\! -\! \eta )}{Q_{n-1}(x)Q_{n-2}(x)}
e^{\eta \p_x}\right )
\left (1\! -\! \frac{Q_n(x\! +\! \eta )
Q_{n-1}(x\! -\! \eta )}{Q_n (x) Q_{n-1}(x)}
e^{\eta \p_x}\right ).}
\end{array}
\eeq

Equation (\ref{d8}) at ${\bf t}=0$ is the difference equation
for $Q_1(x)$:
\beq\label{e5}
\sum_{k=0}^n (-1)^k T^k (x)Q_1(x-(k-1)\eta )=0.
\eeq
The fact that $Q_1(x)$ satisfies this equation is also obvious
from the form of the last operator multiplier in (\ref{e6}).
In a similar way, looking at the last multiplier in (\ref{e6a}),
we conclude that $Q_{n-1}(x)$ satisfies the difference equation
\beq\label{e5a}
\sum_{a=0}^n (-1)^a \frac{T^a(x+(a-1)\eta )}{\phi (x+(a-1)\eta )}\,
\frac{Q_{n-1}(x+(a-1)\eta )}{\phi (x+a\eta )}=0.
\eeq
Equations (\ref{e5}) and (\ref{e5a}) generalize the 
famous Baxter's $TQ$-relation
to $GL(n)$-invariant 
models with $n>2$. Difference equations for the functions
$Q_i(x)$ with $2\leq i\leq n-2$ also exist but have a more
complicated form (see \cite{KLWZ97}).

As it was already mentioned, the fact that $\Psi (x)=
Q_1(x+\eta )=A_1(x)$
is one of solutions to equation ${\bf W}(x,0)\Psi (x)=0$ 
is easily seen from the factorization (\ref{e6}). The
other solutions of this equation are $A_2(x), \ldots , A_n(x)$,
although this is not so easy to see from (\ref{e6}).
Let us show this for $A_2(x)$. To this end, we should change
the order of the points $p_i$:
$$\{p_1, p_2, \ldots , p_n\} \longrightarrow
\{p_2, p_1, \ldots , p_n\}.
$$
We denote the $Q$-functions for this order as $\check Q_i(x)$:
$\check Q_1(x)=A_2(x-\eta )$, $\check Q_i(x)=-Q_i(x)$ for
$i=2, \ldots , n-1$. 
Consider the product of the last two operator multipliers 
in (\ref{e6}). It is easy to check that
$$
\begin{array}{c}
\displaystyle{
\left (1-\frac{Q_{2}(x+\eta )Q_{1}(x-\eta )}{Q_{2}(x)
Q_{1}(x)}e^{-\eta \p_x}\right )
\left (1-\frac{Q_{1}(x+\eta )}{Q_{1}(x)}e^{-\eta \p_x}\right )}
\\ \\
\displaystyle{
=\left (1-\frac{Q_{2}(x+\eta )\check Q_{1}(x-\eta )}{Q_{2}(x)
\check Q_{1}(x)}e^{-\eta \p_x}\right )
\left (1-\frac{\check Q_{1}(x+\eta )}{\check 
Q_{1}(x)}e^{-\eta \p_x}\right ).}
\end{array}
$$
From the right-hand side we see that the operator (\ref{e6}) 
indeed kills the function $A_2(x)$. For the other solutions 
the idea of the proof is similar: to change the order of the
points $p_i$ appropriately, introduce the corresponding functions
$\check Q_i(x)$ and represent the product of several last
factors in (\ref{e6}) in such a way that the rightmost one
would be $\displaystyle{
1-(\check Q_{1}(x+\eta )/\check 
Q_{1}(x))e^{-\eta \p_x}}$.
We omit technical details which 
can be found in \cite{Z98}.

Finally, let us compare the system of Bethe equations (\ref{d13}) 
obtained as a discrete dynamical system for 
zeros of tau-functions 
in the context of the mKP hierarchy with the system
(\ref{eig2}) obtained by the standard methods of the theory of
integrable spin chains (the nested Bethe ansatz).
Although they look quite similarly, there are differences. 
In fact the differences are  
due to the two different (but equivalent) 
approaches to the problem and can be
eliminated merely by a change of notation. Indeed, the standard view
on the nested Bethe ansatz is a gradual ``undressing'' of the
original $GL(n)$-problem which is done in $n-1$ steps by transition
from $GL(n)$ to $GL(n-1)$, then 
from $GL(n-1)$ to $GL(n-2)$ and so on, up to 
the $GL(1)$-model which is trivial. In the mKP-picture this
corresponds to the undressing chain (\ref{chain}). However, 
in the mKP context, it is more natural to invert the arrows 
and follow this chain 
in the opposite direction, from right to left, then each 
step is a ``dressing'' transformation and the chain 
becomes the dressing chain discussed in Section \ref{section:dressing}.
Note that now nothing prevents to continue it infinitely
to the left.
Therefore, to identify (\ref{d13}) and (\ref{eig2}), we should
change the notation in accordance
with this understanding. Namely, after the identification 
\beq\label{id1}
N_m ={\cal N}_{n-m}, \quad 
v_{\alpha}^{(m)}=w_{\alpha}^{(n-m)}, \quad
p_m=g_{n-m+1}
\eeq
equations (\ref{d13}) coincide with (\ref{eig2}).

\section{Connection with the 
classical Ruij\-se\-na\-ars-Schnei\-der model}

We have seen that any eigenvalue $T(x; {\bf t})$ of the master 
$T$-operator 
as a function of the times ${\bf t}$ 
and $t_0=x$ is a solution 
of the mKP hierarchy in the bilinear form (the tau-function),
and this tau-function is a (quasi)polynomial in $x$:
\beq\label{master13}
T(x; {\bf t})=e^{t_1 {\rm tr}\, {\bf g} + t_2 {\rm tr}\, {\bf g}^2 +\ldots}
\prod_{k=1}^{N}\left (x-x_k({\bf t})\right )
\eeq
(the exponential factor is restored from the limit $x\to \infty$).
The roots of this polynomial depend on 
$t_i$.

The connection with the classical 
Ruij\-se\-na\-ars-Schnei\-der system of particles becomes clear
if one addresses dynamics of zeros of $T(x; {\bf t})$ as functions
of the times. 
The dynamics of zeros of polynomial tau-functions is a well known subject
in the theory of integrable nonlinear partial differential equations.
In the works by Krichever and others 
(see \cite{AMM77}--\cite{Z19}) it was found that  
this dynamics is described 
by equations of motion of integrable 
many-body systems of the Calogero-Moser 
and Ruijsenaars-Schneider type. 
In particular, the dynamics of zeros of the tau-function of 
the mKP hierarchy
of the form (\ref{master13}) in the time $t_k$ 
coincides with the dynamics of the Ruijsenaars-Schneider
system of particles \cite{RS86} (which is also known as a relativistic 
deformation of the Calogero-Moser system \cite{Calogero75,Moser75})
with respect to the 
$k$-th Hamiltonian flow. For example, 
the equations of motion in the time $t_1$ have the form
\beq\label{master14}
\ddot x_i=-\sum_{k\neq i}\frac{2\eta^2 
\dot x_i \dot x_k}{(x_i-x_k)((x_i-x_k)^2-\eta^2)},
\eeq
where the dot means the $t_1$-derivative.
The parameter $\eta$ has the meaning of the inverse velocity of light. 
In the limit $\eta \to 0$ one reproduces the Calogero-Moser system of particles.

The Ruijsenaars-Schneider system is 
Hamiltonian with the Hamiltonian function
$$
{\cal H}=\sum_{i=1}^N e^{P_i}
\prod_{k\neq i}\frac{x_i-x_k+\eta}{x_i-x_k}\,, \qquad
\{P_i, x_k\}=\delta_{ik}.
$$
Note that the velocities of the particles are
\beq\label{int4}
\dot x_i =\frac{\p {\cal H}}{\p P_i}=e^{P_i}
\prod_{j\neq i} \frac{x_i-x_j+\eta }{x_i-x_j},
\eeq
so
$\displaystyle{
{\cal H}=\sum_{i=1}^N \dot x_i.}
$

The system is known to be integrable: 
there are $N$ independent conserved 
quantities in involution 
${\cal I}_k$, $k=1, \ldots , N$, and ${\cal I}_1={\cal H}$. 
Explicitly, they are given by the formula
\beq\label{int1}
{\cal I}_k= \sum_{I\subset \{1, \ldots , N\}, \, |I|=k}
\exp \Bigl ( 
\sum_{i\in I}P_i\Bigr ) \prod_{i\in I, j\notin I}
\frac{x_i-x_j+\eta}{x_i-x_j}, \quad k=1, \ldots , N.
\eeq
In terms of the velocities, the integrals of motion read
\beq\label{int5}
{\cal I}_k=\sum_{I\subset \{1, \ldots , N\}, \, |I|=k}
\,\, \Bigl (\prod_{i\in I}\dot x_i \Bigr )\prod_{i<j, \, i,j\in I}
\frac{(x_i-x_j)^2}{(x_i-x_j)^2-\eta^2}.
\eeq

We should recall that the classical $N$-body Ruijsenaars-Schneider model 
admits a commutation representation in the form of the matrix
Lax equation
\beq\label{qc1}
\dot L=[L,M]
\eeq
for $N\! \times \! N$ matrices $L,M$ whose matrix elements are functions of
$x_j$ and $\dot x_j$. The matrix $L$ is called the Lax matrix, its explicit form
is
\beq\label{qc2}
L_{ij}=L_{ij}\Bigl (\{\dot x_l\}_N, \{x_l\}_N\Bigr )=
\frac{\dot x_i}{x_i-x_j-\eta}, \qquad i,j=1, \ldots , N.
\eeq
For our purposes we 
do not need the explicit form of the matrix $M$. 
Equations of motion (\ref{master14}) are equivalent to the matrix equation
(\ref{qc1}). The Lax equation implies that the time evolution 
of the Lax matrix $L(0)\rightarrow L(t)$ 
is an isospectral transformation, i.e.,
eigenvalues of the Lax matrix (and all 
symmetric functions of them) are integrals of motion. 
It is not difficult to see that the characteristic 
polynomial of the Lax matrix is
the generating function of the integrals of motion ${\cal I}_k$:
\beq\label{int9}
\det_{N\times N} (zI-L)=z^N + 
\sum_{k=1}^{N}\eta^{-k}\,
{\cal I}_k z^{N-k},
\eeq
where $I$ is the unity matrix. 
Therefore, the integrals of motion ${\cal I}_k$ are given by
elementary symmetric polynomials 
$e_k(\xi_1, \ldots , \xi_N)$ (see (\ref{he}))
of eigenvalues $\xi_i$ of the Lax matrix:
\beq\label{int10}
{\cal I}_k =(-1)^k \eta^k e_k(\xi_1, \ldots , \xi_N).
\eeq

The first formula in (\ref{master10a}) 
tells us that the eigenvalue $T(x)$ of the
transfer matrix ${\bf T}(x)$ (\ref{master1a}) is
$$
T(x)=\p_{t_1}\log T(x; {\bf t})\Bigr |_{{\bf t}=0}.
$$
Plugging here (\ref{master13}) and comparing with 
(\ref{master2}), we obtain the following important relation
between eigenvalues of the quantum Hamiltonians ${\bf H}_i$
(\ref{master2a}) and initial velocities of the Ruijsenaars-Schneider
particles:
\beq\label{master13a}
\eta H_i=-\dot x_i(0).
\eeq
This relation allows one to make the connection with
the Ruijsenaars-Schneider system more precise. 
For the latter system the standard problem 
of classical mechanics is to determine
time evolution of the $x_i$'s from given initial conditions
$x_i(0), \dot x_i(0)$, which determine values of all
integrals of motion and can be arbitrary. 
However, in applications to quantum spin chains
the problem is posed in a different way.
To understand this, recall that in the Krichever's 
method which was used in Section \ref{section:polynomial}
for construction of polynomial solutions to the mKP hierarchy
only initial coordinates $x_i(0)$ (zeros of the tau-function) 
enter the game as arbitrary
parameters (they are implicitly determined from the $a_{im}$'s
in (\ref{rat2a})), but as soon as the conditions
(\ref{rat2a}) are imposed, the initial velocities 
$\dot x_i(0)$ with respect to $t_1$
can not be arbitrary. Instead, the arbitrary parameters are
the points $p_i$ which are singular points of the spectral curve.
The integrals of motion of the Ruijsenaars-Schneider system
for the $x_i$'s are determined by the spectral curve, so their
values have to be expressed through the $p_i$'s.
Therefore, we see that 
the problem should be posed in the following unusual way:
given $x_i=x_i(0)$ and values of all higher integrals of motion,
to find $\dot x_i(0)$ (which, according to 
(\ref{master13a}), give $H_i$). 
This latter problem has more than one solution,
and each solution corresponds to a common eigenstate of the
quantum transfer matrices of the spin chain.

Taking all this into account, we can expect that 
eigenvalues of the Lax matrix (\ref{qc2}) for the class of solutions 
constructed in Section \ref{section:polynomial} are expressed 
through the given parameters $p_i$ (which on the 
spin chains side are the twist parameters). To establish a precise
relation, we use the differential-difference equation (\ref{diff2})
for the adjoint wave function, i.e.,
\beq\label{diff2a}
-\p_{t_1}\psi^* (x+\eta , {\bf t};z)=
\psi^* (x ,{\bf t};z)+
\p_{t_1}\log \Bigl (\frac{\tau (x+\eta , {\bf t})}{\tau (x, {\bf t})}
\Bigr )\,
\psi^* (x+\eta , {\bf t};z),
\eeq
with $\psi^*$ given by 
(\ref{BA2}) with the tau-function from (\ref{rat6}). 
This function has $N$ simple poles 
at the points $x_i=x_i({\bf t})$.
We can represent it as a sum of simple pole terms:
\beq\label{diff4}
\psi^* (x, {\bf t};z)=z^{-x/\eta}e^{-\xi ({\bf t}, z)}
\left (c_0^*(z)+\sum_{i=1}^N \frac{c_i^*({\bf t}, z)}{x-x_i}\right )
\eeq
with some coefficients $c_i^*$ parametrizing residues at the poles
which can depend on $z$ and ${\bf t}$ but not on $x$. 
Plugging this ansatz into (\ref{diff2a}), we obtain the relation
$$
c_0^* +\sum_i \frac{c_i^*}{x-x_i+\eta}
-z^{-1}\sum_i \frac{\dot c_i^*}{x-x_i+\eta}
-z^{-1}\sum_i \frac{c_i^*\dot x_i}{(x-x_i+\eta )^2}
$$
$$
=c_0^* +\sum_i \frac{c_i^*}{x-x_i} +z^{-1}\sum_j
\left (\frac{\dot x_j}{x-x_j}-\frac{\dot x_j}{x-x_j+\eta}\right )
\left (c_0^* +\sum_i \frac{c_i^*}{x-x_i+\eta}\right )
$$
both sides of which are rational functions of $x$ with simple
poles at $x=x_i$ and second order poles at $x=x_i-\eta$. It is easy
to see that the highest order poles cancel identically and we need 
only to identify the residues at $x=x_i$ and $x=x_i-\eta$.
For our purpose here it is enough to consider the poles at
$x=x_i$. Their cancellation leads to the following system
of linear equations for the coefficients $c_i^*$:
\beq\label{diff5}
zc_i^* +\dot x_i c_0^* +\dot x_i \sum_j \frac{c_j^*}{x_i-x_j+\eta}=0,
\qquad i=1, \ldots , N.
\eeq
It can be written in a matrix form:
\beq\label{diff6}
{\bf c}^* \dot X^{-1}\Bigl (zI-L\Bigr )=-c_0^* {\bf e},
\eeq
where ${\bf c}^* =(c_1^* , \ldots , c_N^*)$ is a row vector,
$\dot X=\mbox{diag}\, (\dot x_1, \ldots , \dot x_N)$,
$L$ is the Lax matrix (\ref{qc2}) and ${\bf e}=(1,1 , \ldots , 1)$.
The solution of this system is
\beq\label{diff7}
{\bf c}^*({\bf t}, z)=-c_0^*(z) {\bf e}\Bigl (zI-L\Bigr )^{-1}
\dot X.
\eeq

It remains to find $c_0^*(z)$. To this end, it is enough to tend
$x\to \infty$ in (\ref{diff4}) and (\ref{BA2}) with the
tau-function given by (\ref{rat6}) and compare the results.
This gives
\beq\label{diff8}
c_0^*(z)=z^n\prod_{i=1}^n (z-p_i)^{-1}.
\eeq
By construction of the adjoint wave 
function (\ref{diff4}) we know that it has multiple poles 
at $z=p_i$, $i=1, \ldots , n$. Therefore, looking at
(\ref{diff7}) with $c_0^*$ given by (\ref{diff8}), 
we conclude that the eigenvalues of the Lax matrix should be 
identified with
$p_1, \ldots , p_n$, with each $p_i$ being 
in general multiple eigenvalue with multiplicity $M_i$
(then the right-hand side of (\ref{diff7}) has a pole of order 
$M_i+1$ at this point, as it should).

Relating this result to quantum spin chains, we can reformulate
it as follows.
Consider the Lax matrix (\ref{qc2}) $L(0)$ with 
the substitution $\dot x_i(0)=
-\eta H_i$, where $H_i$ are eigenvalues (corresponding to
a common eigenstate) of the quantum
Hamiltonians ${\bf H}_i$ of the generalized twisted 
inhomogeneous spin chain given by (\ref{eig3}):
$$
L_{ij}(0)=L_{ij}\Bigl (\{-\eta H_l\}_N, \{x_l\}_N\Bigr )=
\frac{\eta H_i}{x_j-x_i+\eta}.
$$
Then the spectrum of $L$ has the
following specific form:
\beq\label{qc3}
\mbox{Spec}\, L \Bigl (\{-\eta H_i\}_N, \{x_i\}_N\Bigr )=
\Bigl (\underbrace{p_1, \ldots , p_1}_{M_1}, \, \underbrace{p_2, \ldots , p_2}_{M_2}, \,
\ldots , \, \underbrace{p_n, \ldots , p_n}_{M_n}\Bigr ),
\eeq
where $M_a$ are eigenvalues of 
the operators ${\bf M}_a$ (\ref{eig0}) on the eigenstate
of the transfer matrix (we recall that 
$\displaystyle{\sum_{a=1}^n M_a=N}$) and $p_i$ are twist parameters
(elements of the diagonal twist matrix ${\bf g}$). 
In other words, the characteristic polynomial of the Lax matrix
is
\beq\label{qc4a}
\det \Bigl [zI- L \Bigl (\{-\eta H_i\}_N, \{x_i\}_N
\Bigr )\Bigr |_{BE} \Bigr ]=
\prod_{i=1}^n (z-p_i)^{M_i},
\eeq
where $L \Bigl (\{-\eta H_i\}_N, \{x_i\}_N\Bigr )$ 
is taken on a solution to the
Bethe equations. 
A very technical proof of this 
result, which essentially uses the nested Bethe equations (\ref{eig2}),
can be found in \cite{GZZ14}. Here we have suggested another proof,
which is easier and more instructive. 

This result is also known as 
quantum-classical duality for integrable systems 
(quantum spin chain versus classical Ruijsenaars-Schneider). 
The remarkable relation between the two so different systems
survives (and remains nontrivial) 
also in the limit $\eta \to 0$, where on the quantum side we have
the Gaudin model and the 
Calogero-Moser system on the classical side
(see \cite{GZZ14,ALTZ14} for details). 

From a more general viewpoint,
the quantum-classical duality is a remarkable relation
between the joint spectra of commuting quantum Hamiltonians
and intersection of two
Lagrangian submanifolds of the $2N$-dimensional 
phase space of an $N$-body classical 
integrable system of particles.
A Lagrangian submanifold is an 
$N$-dimensional submanifold in the $2N$-dimensional
phase space such that the restriction of the symplectic form 
$\omega =\displaystyle{\sum_{i=1}^{N}dP_i \wedge dx_i}$ to it
is identically equal to zero.
In the relation mentioned above, the first
 Lagrangian submanifold is the $N$-dimensional
 hyperplane corresponding to fixing all coordinates $x_j$ of
the classical particles, while the second one is the level set
 of the $N$ independent integrals of motion in involution.
Their dimensions are
 complimentary, and thus they intersect in a finite number of points.
 The essence of the quantum-classical
 duality is that the values of velocities $\dot x_j$ of the particles
 at the intersection points provide spectra of commuting  
 quantum Hamiltonians of some quantum integrable
 model (one of the examples is the twisted 
 inhomogeneous $GL(n)$-invariant
 spin chain considered in the present paper).
Different intersection points correspond to
different eigenstates of the commuting quantum Hamiltonians.

Let us say a few words about the meaning of this result.
In particular, it makes it possible 
to solve the spectral problem for the quantum 
Hamiltonians without addressing the Bethe ansatz at any step.
Instead, one should solve an ``inverse spectral problem'' 
for the Lax matrix of the classical integrable
system of particles of Ruijsenaars-Schneider or
Calogero-Moser type. Namely, let $\{x_i\}_N$ be inhomogeneity 
parameters of the spin chain and ${\bf g}=
\mbox{diag}\, (p_1, p_2, \ldots , p_n)$ its
twist matrix. Let the eigenvalues of the Lax matrix be equal to the eigenvalues 
$p_i$ of the twist matrix, with some multiplicities $M_i$ such that 
$\displaystyle{\sum_{i=1}^n M_i=N}$. 
This fixes values of all the Ruijsenaars-Schneider
integrals of motion according to (\ref{int10}). 
Then the spectrum of
the non-local spin chain Hamiltonians ${\bf H}_j$ in the sector where eigenvalues of the weight
operators ${\bf M}_i$ are equal to $M_i$ 
is given by the values of $H_j$ such that the
matrix $\displaystyle{L_{ij}=\frac{\eta H_i}{x_j-x_i+\eta}}$ 
has the prescribed spectrum (\ref{qc3}). 
This kind of duality suggests 
an alternative way to calculate joint spectra of commuting
quantum transfer matrices without use of the coordinate or 
algebraic Bethe ansatz technique. 
There is also no need in such an unavoidable
intermediate step as solving the 
Bethe equations. The spectra of {\it quantum} Hamiltonians
appear to be encoded in algebraic properties of the Lax matrix for a very different {\it purely classical} model. 

To be more precise, combining (\ref{int5}), (\ref{int10}) and (\ref{master13a}),
we obtain a system of algebraic equations for the joint spectrum
of the Hamiltonians ${\bf H}_i$:
 \beq\label{int5a}
\sum_{I\subset \{1, \ldots , N\}, \, |I|=k}
\,\, \Bigl (\prod_{i\in I}H_i \Bigr )\prod_{i<j, \, i,j\in I}
\frac{(x_i-x_j)^2}{(x_i-x_j)^2-\eta^2}=
e_k(\{\xi_i\}_N),
\eeq
where $e_k$ are the elementary
symmetric polynomials. 
For example, 
$$
e_1(\{\xi_i\}_N)=\sum_i\xi_i, \quad
e_2(\{\xi_i\}_N)=\sum_{i<j}\xi_i \xi_j, \quad \mbox{and so on}.$$ 
The set $\{\xi_1, \ldots , \xi_N\}$ consists of $N$ 
given eigenvalues of the Lax matrix which are the twist parameters
$p_i$ with multiplicities. In contrast to the Bethe equations,
which are equations for some auxiliary variables,
(\ref{int5a}) are equations for the spectrum itself. 

Finally, let us present the formulas which express the wave function,
and its adjoint for polynomial solutions of the mKP hierarchy
through the Lax matrix $L$ (\ref{qc2}). We give them here without
derivation (see \cite{TZZ15,Iliev07} for details):
\beq\label{psi}
\begin{array}{l}
\displaystyle{
\psi (x, {\bf t}; z)=\prod_{k=1}^n (1-p_kz^{-1})\,
z^{x/\eta}e^{\xi ({\bf t}, z)}
\frac{\det \Bigl [ (xI-X)(zI-L)-\eta L\Bigr ]}{\det 
\Bigl (xI-X\Bigr )\Bigl (zI-L\Bigr )},}
\\ \\
\displaystyle{
\psi^* (x, {\bf t}; z)=\prod_{k=1}^n (1-p_kz^{-1})^{-1}
z^{-x/\eta}e^{-\xi ({\bf t}, z)}
\frac{\det \Bigl [ (zI-L)(xI-X)+\eta L\Bigr ]}{\det 
\Bigl (xI-X\Bigr )\Bigl (zI-L\Bigr )},}
\end{array}
\eeq
where $X=X({\bf t})=\mbox{diag}(x_1({\bf t}), 
\ldots , x_N({\bf t}))$. Using (\ref{BA1}),
(\ref{BA2}), one can see from (\ref{psi}) 
that the tau-function for 
this class of solutions is given by
the following determinant formula:
\beq\label{tau}
\tau (x, {\bf t})=\prod_{i=1}^n \Bigl (p_i^{x/\eta}
e^{\xi ({\bf t}, p_i)}\Bigr )
\, \det_{N\times N} \left (
xI-X_0 +\eta \sum_{k\geq 1}kt_k L_0^k\right ),
\eeq
where $X_0 =X(0)$, $L_0=L(0)$.

\section{Concluding remarks}

We have reviewed the approach to quantum integrable
models solvable by Bethe ansatz 
developed in \cite{AKLTZ13}--\cite{Z15}, refining some arguments 
from these works and making the results more detailed. The essence 
of our approach is diagonalization of quantum transfer matrices
by methods of the classical soliton theory, avoiding the Bethe
ansatz procedure. It seems to us that 
the translation from the language of quantum 
integrability to the one of classical integrable hierarchies
is suggestive and instructive and adds something important 
to the deeper understanding of both areas of mathematical physics. 
Objects and notions from the arsenal of the algebraic 
Bethe ansatz find their natural counterparts 
in the classical theory of the soliton
equations. 
For example,
factorization of the non-commutative generating function of
transfer matrices for fundamental representations, which is a key
step of the algebraic Bethe ansatz solution, on the classical side 
manifests itself
as factorization of the difference wave operator of order $n$
into product of $n$ first order difference operators, each of them,
being applied to the wave 
function, produces a B\"acklund transformation. 

As is shown in \cite{TZZ15}, this approach, as well as the 
quantum-classical duality, can be extended to
supersymmetric $GL(n|m)$-invariant 
spin chains (graded magnets) with rational $R$-matrices.
However, the connection with the mKP hierarchy needs to be
made more precise. Presumably, the wave operators relevant
to $GL(n|m)$-invariant models are no longer finite difference
operators but are of the form
${\bf W}(x)={\bf W}_1(x){\bf W}^{-1}_2(x)$, where 
${\bf W}_1$, ${\bf W}_2$ are difference operators of
orders $n$ and $m$.

In this paper we were restricted to quantum models with $R$-matrices
that are rational functions of the spectral parameter.
As we have seen,
on the classical side they correspond to rational 
solutions of the integrable hierarchy, for which the
tau-function is a polynomial or quasipolynomial. As it was 
argued in \cite{Z13,BLZZ16}, this approach can be extended,
mutatis mutandis, to spin chains
with trigonometric $R$-matrices (of the XXZ type), which are 
related to quantum deformations $U_q(gl_n)$ of the universal
enveloping algebras with a deformation 
parameter $q=e^{\gamma}$. (In these works it was assumed, however, 
that $q$ is not a root of unity.)
On the classical side, such spin chains correspond to mKP
tau-functions which are trigonometric polynomials of $x$
(i.e., Laurent polynomials of $e^{\gamma x}$). 
The solutions of this class, too, can be characterized by 
Krichever's conditions. However, instead of (\ref{rat2a})
they have the form
\beq\label{c1}
\sum_{m=-M_i/2}^{M_i/2} a_{im}\psi (x, {\bf t}, p_ie^{2\gamma m})=0,
\eeq
where the sum goes over all integer numbers between $-M_i/2$
and $M_i/2$ for even $M_i$ and over all half-integer numbers
between $-M_i/2$ and $M_i/2$ for odd $M_i$. The poles of
the adjoint wave function of high 
orders $M_i+1$ at the points $p_i$ 
in the trigonometric case 
become ``strings'' of simple poles at the $M_i+1$ points
$p_ie^{-2\gamma M_i}, p_ie^{-2\gamma (M_i-1)}, \ldots ,
p_ie^{2\gamma M_i}$. The case when $q$ is a root of unity is
more complicated and requires a separate investigation.
We hope to revisit the trigonometric 
case in a separate publication.

As far as a
possibility to extend this approach to quantum 
models with elliptic
$R$-matrices is concerned, it 
still remains to be a challenging open problem.
On the first glance it seems that in this most general case 
the master $T$-operator can be defined by the same formula 
(\ref{master10}).
However, one can see that
the transfer matrices ${\sf T}_{\lambda}(x)$ with 
different $\lambda$'s have 
different monodromy properties under shifts by the periods.
This means that the master $T$-operator defined by
equation (\ref{master10}), being a linear combination
of them, fails to be an ``elliptic polynomial'' of $x$, and so
the existing theory of elliptic solutions to the mKP hierarchy
is not applicable.

Finally, let us recall that the quantum-classical duality
receives a formal ``explanation'' if one treats it as 
a limiting case 
of the so-called Matsuo-Cherednik correspondence, which 
was established in \cite{Matsuo92,Cherednik94} 
(see also \cite{ZZ17,ZZ17a}, where it was generalized and 
connected with the quantum-classical duality). The Matsuo-Cherednik
correspondence, sometimes called the quantum-quantum duality, 
connects solutions 
to the (quantum) Knizhnik-Zamolodchikov equations (which can be 
regarded as a non-stationary extension of the spectral 
problem for the spin chain 
Hamiltonians ${\bf H}_j$) with 
stationary wave functions of the quantized 
Ruijsenaars-Schneider model. The parameter that controls 
non-stationarity in the former plays the role of the 
Planck's constant in the latter. Tending it to zero makes 
the former problems stationary but still quantum, while 
the latter models become classical, and this is the way
how the quantum-quantum (Matsuo-Cherednik) duality turns 
to the quantum-classical duality. 
As we have seen, the quantum-classical duality
can be lifted to the level
of integrable hierarchies of nonlinear equations like KP or mKP. 
It is then natural to ask whether something similar is true 
for the quantum-quantum duality, i.e., could it be obtained 
from a more general connection with an integrable hierarchy 
of KP or mKP type or its quantized version.

\section*{Acknowledgments}

I am grateful to my co-authors 
A. Alexandrov, M. Beketov, A. Gorsky,
V. Kazakov, I. Krichever, S. Leurent, A. Liashyk, O. Lipan,
A. Sorin, Z. Tsuboi, P. Wiegmann, A. Zotov for
collaboration in the works 
\cite{KLWZ97,KSZ08}, \cite{AKLTZ13}--\cite{BLZZ16}, 
to N. Reshetikhin and A. Veselov for discussions 
and to J. Harnad for careful reading
of the manuscript and valuable remarks.
Also, I would 
like to thank organizers of the Beijing Summer Workshop in 
Mathematics and Mathematical Physics at BIMSA, China
(June 24 -- July 5, 2024) for hospitality. 
This work has been supported in part 
within the state assignment of NRC 
``Kurchatov institute'' (sections 2--4).

\end{document}